\documentclass[12pt,tightenlines,eqsecnum,floats,aps,amsmath,amssymb,nofootinbib,prd]{revtex4}

\usepackage{setspace}
\usepackage{subfig}
\usepackage{amsmath,amssymb,amsfonts,amsthm,mathrsfs}
\usepackage{graphicx}
\usepackage{enumerate}

\def\be{\begin{equation}}
\def\ee{\end{equation}}
\def\ba{\begin{eqnarray}}
\def\ea{\end{eqnarray}}
\def\bi{\begin{itemize}}
\def\ei{\end{itemize}}

\begin{document}

\title{Gravitation in terms of observables}

\author{Rodolfo Gambini$^1$ and Jorge Pullin$^2$}
\affiliation{1. Instituto de Física, Facultad de Ciencias, Iguá 4225, esq. Mataojo,
11400 Montevideo, Uruguay. \\
 2. Department of Physics and Astronomy, Louisiana State University,
Baton Rouge, LA 70803-4001.}

\begin{abstract}
  In the 1960's, Mandelstam proposed a new approach to gauge theories
  and gravity based on loops. The program was completed for
  Yang--Mills theories by Gambini and Trias in the 1980's. In this
  approach, gauge theories could be understood as representations of
  certain group: the group of loops. The same formalism could not be
  implemented at that time for the gravitational case. Here we would
  like to propose an extension to the case of gravity. The resulting
  theory is described in terms of loops and open paths and can provide
  the underpinning for a new quantum representation for gravity
  distinct from the one used in loop quantum gravity or string
  theory. In it, space-time points are emergent entities that would
  only have quasi-classical status. The formulation may be given
  entirely in terms of Dirac observables that form a set of
  gauge invariant functions that completely define the Riemannian
  geometry of the spacetime. At the quantum level this formulation
  will lead to a reduced phase space quantization free of any
  constraints.

\end{abstract}
\maketitle

\section{Introduction}

There exists a renewed interest in the description in terms of
observables of gauge theories and gravity. Recently, Giddings and
Donnelly \cite{GiDo} proposed explicit constructions that extend the
observables associated to gauge theories to the case of gravitation
for weak fields. They note that an important feature of the resulting
quantum theory of gravity is the algebra of observables, that becomes
non-local. Observable-based techniques are also used in several modern
developments attempting to extract information from quantum gauge
theories \cite{nonlocal}.  The most ambitious attempt to describe
gravity intrinsically without coordinates was proposed by Mandelstam
in the 1960's \cite{mandelstam}. The approach did not flourish because
the intrinsic description loses completely the notion of space-time
point, and becomes difficult to recover it even classically. Paths
that end in the same physical point in this description cannot be
easily recognized. In the 1980's Gambini and Trias \cite{gatr} showed
that gauge theories arise as representations of the group of loops in
certain Lie groups. The complete geometric structure of gauge theories
can be recovered from identities obeyed by the infinitesimal
generators of the group of loops. The possibility of extending this
description to the gravitational case did not appear possible due to
the issues we mentioned with Mandelstam's approach. In this paper we
will show how to extend the notion of the group of loops and its
representations which arise in gauge theories to the gravitational
case. This leads to a complete classical description of gravitation
without coordinates.  The metric is everywhere referred to local
frames parallel transported starting from a given point. In such
frames it takes the Minkowskian form. The geometrical content of the
theory is completely recovered by relations between reference frames
obtained by parallel transport along paths that differ by an
infinitesimal loop and is given by the Riemann tensor.  Although the
construction is based on loops, it differs from the one underlying the
usual loop representation of gauge theories and gravity. In the loop
representation the objects constructed are gauge invariant whereas in
the present construction the objects are both gauge invariant and
space-time diffeomorphism invariant. That is, the objects are Dirac
observables.  This leads to a theory that {\em does not involve
  diffeomorphisms} and may allow to bypass at the quantum level the
LOST-F \cite{lostf} theorem that leads to a discrete structure in the
Hilbert space of ordinary loop quantum gravity and conflicts with the
differentiability of the group of loops. The latter is crucial to
recover the kinematics of gauge theories and gravity in this context.

The organization of this paper is as follows: In section II we make a
brief review of the group of loops on differential manifolds. In
section III we introduce gauge theories as representations of the
group of loops. In section IV we recall the Mandelstam approach, in
terms of intrinsic paths, to gravity and discuss some of its
problems. In section V we extend the loop techniques to intrinsic
paths. In section VI we show that an intrinsic description of gravity
arises as a representation of the group of loops in the Lorentz group.
In section VII we establish the relation between the intrinsic and
coordinate descriptions. In section VIII we show that the intrinsic
and coordinate representations of gravity are equivalent at the
classical level but they are not equivalent at the quantum level. In
section IX we present an intrinsic path dependent Lagrangian formalism
for arbitrary path dependent fields. In section X we analyze the
relation between path dependence and diffeomorphisms. In section XI we
show how to extend the Hamiltonian techniques to intrinsic
paths. Finally in section XII we present some concluding remarks.

\section{The group of loops: a brief review}

\subsection{Holonomies and the definition of loops}
We will briefly review some notions of the group of loops. For a more
extensive treatment see \cite{gapubook,gatr}.

     We start with a set of parametrized 
curves on a manifold $M$. We assume they are continuous and  piecewise
smooth. There is no real need to have the curves parameterized but we
do it to fix ideas. A curve $p$ is a map
\begin{equation}
p:[0,s_{1}]\cup[s_{1},s_{2}]\cdots[s_{n-1},1]
\;\rightarrow\;M
\end{equation}
that is smooth in each closed interval $[s_{i},s_{i+1}]$ and continuous in
the whole domain.
Given  two
piecewise smooth  curves  $p_{1}$ and $p_{2}$ where the end 
point of $p_{1}\;$is the same as the beginning point of $p_{2}$,
the composition curve $p_{1}\circ p_{2}$ is given by:
\begin{equation}
p_{1}\circ p_{2}(s)=\left\{ \begin{array}{ll}
                  p_{1}(2s),  & \mbox{for$\;s\in [0,1/2]$} \\
                  p_{2}(2(s-1/2))& \mbox{for$\;s\in [1/2,1]$}.
     \end{array}
     \right.
\end{equation}

     The curve traversed in the opposite orientation (``opposite
curve'') is given by
\begin{equation}
p^{-1}(s):=p(1-s).
\end{equation}
     We also consider closed curves $l,m,...$,
that is, curves which start and end
at the same point $o$.
We call
 $L_{o}$ the set of all such closed curves. The set $L_{o}$ is a
semi-group under the  composition  law $(l,m)  \rightarrow  l\circ  m$.
The  identity  element (``null curve'')
is  defined  to  be  the  constant  curve
$i(s)=o$ for any $s$ and any parametrization. However,  we  do  not
have a group structure, since the opposite curve $l^{-1}$  is  not  a
group inverse in the sense that $l\circ l^{-1}\neq i$.

     Holonomies are given by the parallel transport around closed
curves.  The parallel transport around a closed curve $l\in L_{o}$ is a
map from the fiber over $o$ to itself given by the path ordered
exponential,
\begin{equation}
H_{A}(l)=P \exp\int_{l}A_{a}(y)dy^{a}.
\end{equation}\index{path ordered}

The holonomy $H_{A}$ is an element of the group $G$ and the product  denotes
the right action of $G$. The main property of $H_{A}$ is
\begin{equation}
H_{A}(l\circ m)=H_{A}(l)H_{A}(m).
\end{equation}
A change in the choice of the point on the fiber over $o$ 
from $o$ to $o'$ induces the transformation
\begin{equation}
H'_{A}(l)=g^{-1}H_{A}(l)g,
\end{equation}
where $g$ is the holonomy of a path joining $o$ to $o'$.

     In order to transform the set $L_{o}$ into a group, we need to
introduce a further equivalence relation, the idea is to identify all
curves yielding the same holonomy.  These  equivalence classes 
we will from now on call {\em loops}.  We will
denote them with Greek letters, to distinguish them from the
individual curves of the equivalence classes.  Several
definitions of this equivalence relation have been proposed. The
simplest one is that the curves yield the same holonomy for any
connection. Related to it is that two curves that differ by a retraced
path (``tree'') are equivalent since retraced paths (paths that go out
and back along the same curve) do not contribute to the
holonomy. There are other possible definitions but we will not discuss
them here (see \cite{gapubook} and \cite{tavares,lewandowski} for details).

With any of the definitions one can show that the composition
between loops is well defined and is again a loop. In other words if $
\alpha \equiv [l]$ and $ \beta \equiv [m]$ then $ \alpha \circ \beta
=[l \circ m]$ where by $[]$ we denote the equivalence classes. 

\begin{figure}[t]
\includegraphics[height=5cm]{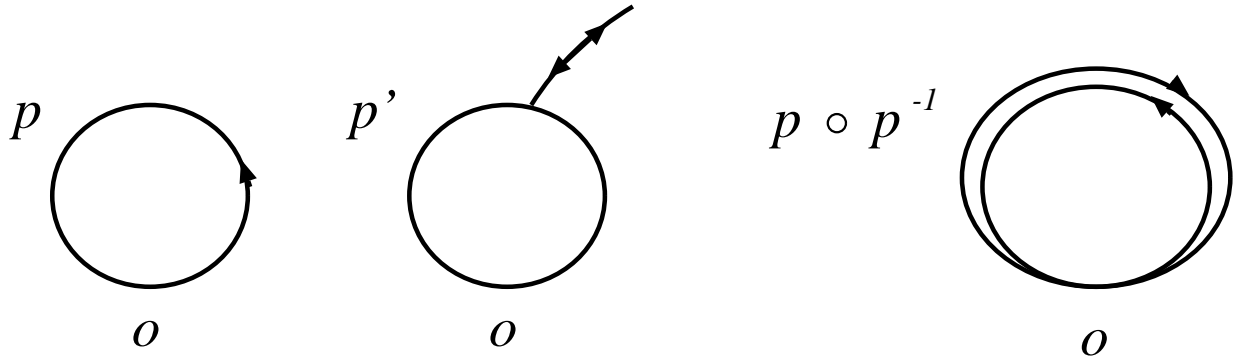}
\caption{Curves $p$ and $p'$ differ by a tree. The composition of a curve
and its inverse is a tree.}
\label{tree}
\end{figure}

        With the equivalence relation defined, it makes
sense to define an inverse of a loop.  Since the composition of a curve with
its opposite yields a tree (see figure \ref{tree}) it is natural,
given a loop $\alpha$, to define
its inverse $\alpha^{-1}$ by $\alpha \circ \alpha^{-1} =
\iota$ where  $\iota$ is the set of closed curves equivalent
to the null curve (thin loops or trees). $\alpha^{-1}$ is the set of
curves opposite to the elements of $\alpha$. We will also denote
inverse loops with an overbar $\alpha^{-1}\equiv \overline{\alpha}$. 

        We will denote the set of loops base-pointed at $o$ by ${\cal
L}_{o}$. Under the composition law given by $\circ$ this set is a
non-Abelian group, which is called the group of loops. 

We have relations between holonomies of composed loops\begin{equation}
H(\alpha \circ \beta) = H(\alpha) H(\beta),
\end{equation}
and that inverses are mapped to each other,
\begin{equation}
H(\alpha^{-1}) = (H(\alpha))^{-1}.
\end{equation}

We will define a set of differential
operators acting on functions of loops that are related to the
infinitesimal generators of the group of loops: the loop and
connection derivatives.

\subsection{The loop derivative}
\index{derivative!loop}\index{loop derivative|see{derivative!loop}}

\begin{figure}[t]
\vspace{-3cm}
\includegraphics[height=20cm]{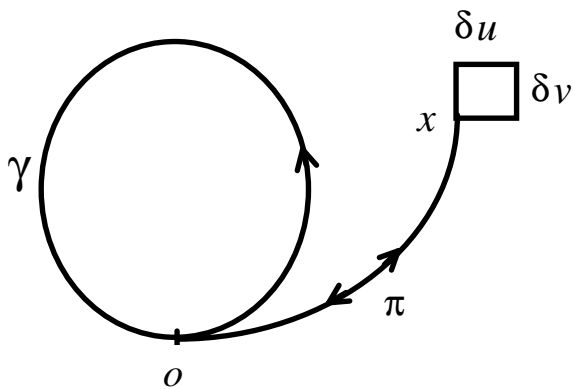}
\vspace{-12cm}
\caption{The infinitesimal loop that defines the loop derivative.}
\label{areadfig}
\end{figure}

\label{loop derivative}
        Given $\Psi(\gamma)$ a continuous, complex-valued  function
of ${\cal L}_o$ we want to consider its variation when the loop
$\gamma$ is  changed by the addition of an infinitesimal loop $\delta
\gamma$ base-pointed at a point $x$ connected by a path $\pi_o^x$ to
the base-point of $\gamma$, as shown in
figure \ref{areadfig}.
That is, we want to evaluate the change in the function when changing
its argument from $\gamma$ to
$\pi_o^x \circ \delta \gamma \circ \pi_x^o \circ \gamma$.  In order to
do this we will consider a two-parameter family of infinitesimal loops
$\delta \gamma$.  Notice that no matter what path $\pi$ one chooses,
the added path is infinitesimal due to the invariance of loops under
re-tracings ---additions of trees--- and therefore induces an
infinitesimal deformation of $\gamma$.  Since spacetimes look flat at
sufficiently small regions, $\delta \gamma$ may be described in a
particular coordinate chart by the curve obtained by traversing the
vector $u^a$ from $x^a$ to $x^a + \epsilon_1 u^a$, the vector $v^a$
from $x^a+\epsilon_1 u^a$ to $x^a+\epsilon_1 u^a +\epsilon_2 v^a$, the
vector $-u^a$ from $x^a+ \epsilon_1 u^a + \epsilon_2 v^a$ to
$x^a+\epsilon_2 v^a$ and the vector $-v^a$ from $x^a+\epsilon_2 v^a$
back to $x^a$ as shown in figure \ref{areadfig}. We will denote these
kinds of curves with the notation
$\delta u \delta v \overline{\delta u} \,\overline{\delta v}$.

For a given $\pi$ and $\gamma$ a loop differentiable 
function depends only on the
infinitesimal vectors $\epsilon_1 u^a$ and $\epsilon_2 v^a$.  We will
assume it has the following expansion with respect to them,
\begin{eqnarray}
\label{exploopd}
\Psi(\pi_o^x\circ \delta \gamma \circ \pi_x^o \circ\gamma)=&&
\Psi(\gamma) + \epsilon_1 u^a Q_a(\pi_o^x) \Psi(\gamma) +
\epsilon_2 v^a P_a(\pi_o^x) \Psi(\gamma)\nonumber\\
&&+{\textstyle {1 \over 2}} \epsilon_1 \epsilon_2
(u^a v^b + v^a u^b) S_{ab}(\pi_o^x) \Psi(\gamma)\nonumber\\
&&+{\textstyle {1 \over 2}} \epsilon_1 \epsilon_2
(u^a v^b - v^a u^b) \Delta_{ab}(\pi_o^x) \Psi(\gamma).
\end{eqnarray}
where $Q,P,S,\Delta$ are differential operators on the space of
functions $\Psi(\gamma)$. If $\epsilon_1$ or $\epsilon_2$ vanishes or
if $u$ is collinear with $v$ then $\delta \gamma$ is a tree and all
the terms of the right-hand side except the first one must
vanish. This means that $Q=P=S=0$. Since the antisymmetric combination
$(u^a v^b - v^a u^b)$ vanishes in any of these cases, $\Delta$ need
not be zero. That is, a function is loop differentiable if for any
path $\pi_o^x$ and vectors $u,v$, the effect of an infinitesimal
deformation is completely contained in the path dependent
antisymmetric operator $\Delta_{ab}(\pi_o^x)$,
\begin{equation}
\Psi(\pi_o^x\circ \delta \gamma \circ \pi_x^o \circ\gamma)=
(1 + {\textstyle {1 \over 2}} \sigma^{ab}(x) \Delta_{ab}(\pi_o^x)) 
\Psi(\gamma),
\label{defloopd}
\end{equation}
where $\sigma^{ab}(x) = 2 \epsilon_1 \epsilon_2 (u^{[a} v^{b]})$ is
the element of area of the infinitesimal loop $\delta \gamma $. 
We will call this operator the loop derivative.

Loop derivatives do not commute. One can show that, 
\begin{equation}
[\Delta_{ab}(\pi_o^x), \Delta_{cd}(\chi_o^y)] = \Delta_{cd}(\chi_o^y)
[ \Delta_{ab}(\pi_o^x)],
\label{conmutad}
\end{equation}
where we have introduced in the right hand side the loop derivative of
functions of 
open paths from which it is immediate to show that
\begin{equation}
\Delta_{ab}(\pi_o^x)[\Delta_{cd}(\chi_o^y)] =
-\Delta_{cd}(\chi_o^y) [\Delta_{ab}(\pi_o^x)].
\end{equation}

Given a function of an open path $\Psi(\pi_o^x)$, a local coordinate chart
at the point $x$ and  a vector in that chart $u^a$, we define the
Mandelstam derivative by considering the change in the function when the
path is extended from $x$ to $x+\epsilon u$ by the infinitesimal
path  $\delta u$ shown in figure 3 as
\begin{equation}
\Psi(\pi_o^x \circ \delta u) = (1 + \epsilon u^a D_a) \Psi(\pi_o^x).
\end{equation}
\begin{figure}[t]
\vspace{-4cm}
\includegraphics[height=20cm]{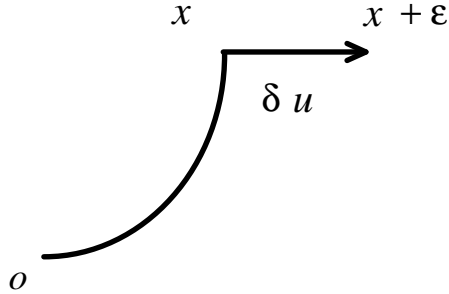}
\vspace{-12cm}
\caption{The extended path defining the Mandelstam derivative,
  $\pi_E=\pi_o^x\circ \delta u$}
\label{mandelder}
\end{figure}

One can derive a Bianchi identity, based on the 
 fundamental idea that ``the boundary of a
boundary vanishes'' and constructing a tree that run the along the
edges of a parallelepiped (see
ref. \cite{gatr}) . The
result is,
\begin{equation}
 D_a \Delta_{bc}(\pi_o^x) +D_b \Delta_{ca}(\pi_o^x) +
D_c \Delta_{ab}(\pi_o^x) =0.
\label{bianchid}
\end{equation}

There is also a Ricci identity,

\begin{equation}
\label{ricciid}
[D_a, D_b] \Psi(\pi_o^x) = \Delta_{ab}(\pi_o^x) \Psi(\pi_o^x).
\end{equation}
This is the analogue of the usual commutator of covariant derivatives
and its relation to the Yang--Mills curvature.
\bigskip

\subsection{The connection derivative}
One can introduce a differential operator
with properties similar to those of the connection or vector
potential of a gauge theory, this allows for a better contact with
the usual formulation of gauge theories.

\begin{figure}
\includegraphics[height=7cm]{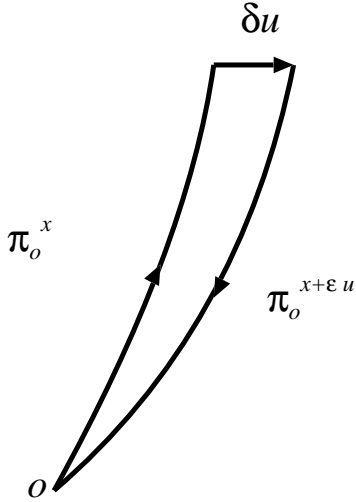}
\caption{The path that
defines the connection derivative. We assume that the point $o$ is in
the same coordinate patch as $x$.} 
\label{condef1} 
\end{figure}

Let us consider a covering of the manifold with overlapping coordinate
patches.  We attach to each coordinate patch ${\cal P}^{i}$ a path
$\pi_o^{y^i_0}$ going from the origin of the loop to a point $y^i_0$
in ${\cal P}^{i}$.  We also introduce a continuous function with
support on the points of the chart ${\cal P}^i$ such that it
associates to each point $x$ on the patch a path $\pi_{y^i_0}^x$.
Given a vector $u$ at $x$, the connection derivative of a continuous
function of a loop $\Psi(\gamma)$ will be obtained by considering the
deformation of the loop given by the path $\pi_o^{y^i_0} \circ
\pi_{y^i_0}^x\circ \delta u \circ
\pi_{x+\epsilon u}^{y^i_0} \circ \pi_{y^i_0}^o$ shown in figure
\ref{condef1}.  
The path
$\delta u$ goes from $x$ to $x+\epsilon u$.  We will say that the
connection derivative $\delta_a$ exists and is well defined if the loop
dependent function of the deformed loop admits an expansion in terms of
$\epsilon u^a$ given by
\begin{equation} 
\Psi(\pi_o^{x} \circ \delta u \circ
\pi_{x+\epsilon u}^o \circ \gamma)=(1+\epsilon u^a \delta_a(x))
\Psi(\gamma),
\end{equation} 
where we have written $\pi_o^x$ to denote the path
$\pi_o^{y^i_0} \circ \pi_{y^i_0}^x$ and similarly for its inverse.

One can show the following relation between the connection and the
loop derivatives,
\begin{equation}
\label{Ddd}
\Delta_{ab}(\pi_o^x) = \partial_{a} \delta_{b}(x)-\partial_{b}
\delta_{a}(x)+ [\delta_a(x),\delta_b(x)],
\end{equation}
again reminiscent of expressions in ordinary Yang--Mills theory.
The 
loop derivative defined by  (\ref{Ddd})
automatically satisfies the Bianchi identities.

The usual relation between connections and holonomies in a local
chart in a gauge theory can also be written in this language, it is given 
by the path ordered exponential,
\begin{equation}
\label{willy}
U(\gamma_0) = {\rm P}\exp\left(\int_{\gamma_0} dy^a \delta_a(y)\right),
\end{equation}
where $U(\gamma_0) \Psi(\gamma)=\Psi(\gamma_0 \circ \gamma)$.
This again is reminiscent of the familiar expression for gauge
theories, which yields the holonomy in terms of the path ordered
exponential of a connection. Through a second path ordered integral it
could be expressed in terms of the loop derivative, embodying the
usual non-Abelian Stokes theorem and illustrated in figure 5.

\begin{figure}[t]
\vspace{0.1 cm}
\includegraphics[height=17cm]{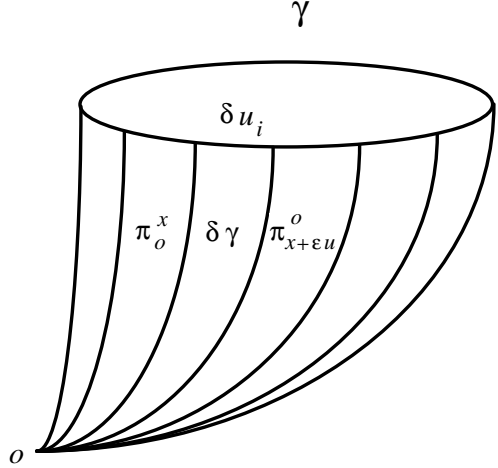}
\vspace{-10.5cm}
\caption{Generating a finite loop using the infinitesimal 
generators combining (2.18) and (2.19).}
\label{construct}
\end{figure}

The relation between the connection and the loop derivative can be
derived in the following way. Consider a deformation going from $\pi^x$ to $\pi^{x+\epsilon}$ given by the displacement vector
field along the path $\pi$ defined as follows: 
Let $\pi^x$ be given by $x^\alpha(\lambda)$ such that
$x^\alpha(\lambda_f)=x^\alpha$ end point of $\pi$, and
$\pi^{x+\epsilon u}$ be
given by $x'{}^\alpha(\lambda)$ such that
$x'{}^\alpha(\lambda_f)=x^\alpha+\epsilon^\alpha$. Then the displacement field
connecting both paths will be given by
$x'^\alpha(\lambda)=x^\alpha(\lambda)+\epsilon^\beta
w_\beta^\alpha(\lambda)$ for all $\lambda$ 
belonging to $[0,\lambda_f]$ and
$w^\alpha_\beta(\lambda_f)=\delta^\alpha_\beta$. From this relation and the definition of
the derivatives we get
\begin{equation}
\delta_\mu(\pi^x)=\int_0^{\lambda_f}\Delta_{\alpha
  \beta}(\pi^x(\lambda)) \dot{x}^\alpha(\lambda) w_\mu^\beta(\lambda)d \lambda.
\end{equation}

Once one attaches to each point of an open region in the manifold a
given path $\pi_o^x$, the connection derivative is an ordinary function
$\delta_\mu(x)=\delta_\mu(\pi_o^x)$. The substitution of (2.19) for the
family of paths $\pi_o^x$ into (2.18) embodies the general form of the non
Abelian Stokes' theorem allowing to write an arbitrary loop deformation
as a ``surface'' integral of loop derivatives. One may therefore
consider the loop derivatives as the infinitesimal generators of the
group of loops.

\section{Kinematics of 
Yang-Mills theories as representations of the group of loops}
\label{rep}
\index{group of loops!representations}

We would like to recall how the kinematical structure of gauge theories emerges
from the group of loops. We consider a map of the group of loops onto
some gauge group $G$,
\begin{equation}
{\cal H}: {\cal L}_0 \rightarrow G, 
\end{equation}
i.e.,
\begin{equation}
\gamma \longrightarrow H(\gamma),
\end{equation}
such that $H(\gamma_1) H(\gamma_2) = H(\gamma_1\circ\gamma_2)$.

Let us consider a specific Lie group, for instance
$SU(N)$, with $N^2-1$ generators  $X^{i}$ such that ${\rm Tr} X^i=0$
and
\begin{equation}
[X^i,X^j] =  C_k^{ij} X^k,
\end{equation}
where $C_k^{ij}$ are the group's structure constants.

Let us compute the action of the connection derivative in this
representation. We use the same prescriptions as in the previous
section
\begin{equation}
(1+\epsilon u^a \delta_a(x)) H(\gamma) = H(\pi_o^x \circ \delta u
\circ \pi_{x+\epsilon u}^o \circ \gamma) = H(\pi_o^x \circ \delta u
\circ \pi_{x+\epsilon u}^o) H(\gamma).
\end{equation}
Since the loop $\pi_o^x \circ \delta u \circ \pi_{x+\epsilon u}^o$ is
close to the identity loop (with the topology of loop space) and
since {\it H} is a continuous, differentiable representation,
\begin{equation}
H(\pi_o^x \circ \delta u \circ \pi_{x+\epsilon u}^o) = 1 + i \epsilon
u^a A_a(x),
\end{equation}
where $A_a(x)$ is an element of the algebra of the group, in our
example of $SU(N)$. That is, $A_a(x)=A_a^i(x) X^i$. Therefore, we see
that through the action of the connection derivative,
\begin{equation}
\label{deltaA}
\delta_a(x) H(\gamma) = iA_a(x) H(\gamma).
\end{equation}

Following similar steps one obtains the action of the loop derivative,
\begin{equation}
\Delta_{ab}(\pi_o^x) H(\gamma) = i F_{ab}(x) H(\gamma),
\end{equation}
where $F_{ab}$ is an algebra-valued antisymmetric tensor field.

{}From equation (\ref{Ddd}) we immediately get the usual relation
defining the curvature in terms of the potential,
\begin{equation}
F_{ab}(x) = \partial_a A_b(x) -\partial_b A_a(x) +i [A_a,A_b].
\end{equation}

From (2.18) and (3.5) we also have that, 
\begin{equation}
H(\eta) = {\rm P} \exp\left(i \oint_\eta dy^a A_a(y)\right),
\end{equation}
yielding the usual expression for the holonomy of the connection
$A_a$. 

In this framework, matter fields can be included considering open
paths. For more details see \cite{gapubook}.

Finally, the usual form of the Ricci identity,
\begin{equation}
[D_a,D_a] = i F_{ab},
\end{equation}
can be obtained directly from the previous expressions, in particular
(\ref{ricciid}). 

This construction allows to recover any gauge theory
with local symmetries associated to a fiber bundle structure. The
extension of this construction to gravity is not trivial. In the
language of fiber bundles it requires the introduction of a soldering
form connecting the fiber to the manifold \cite{hehl}. This is not the 
approach we will take in this paper. In the forthcoming sections
we will develop a formalism that exploits the properties of the group
of loops to construct an intrinsic description of the Riemannian
geometry.

\section{Brief review of Mandelstam's 1962 proposal for quantizing the
gravitational field}

Mandelstam starts with a critique of the usual approaches to
quantizing the gravitational field, which consider c-number
coordinates and q-number metrics and distances. The diffeomorphism
invariance of a theory of quantities like the distances that are
partially quantized through the metric could be problematic. He is
interested in formulating an approach that is coordinate independent
and therefore only framed in terms of q-number physical quantities
associated to intrinsically defined paths without any ambiguity
associated with coordinate conditions, and all distances that appear
in the theory will be physical distances. He focuses on paths in
space-time (manifold plus metric) constructed by starting from a
reference point (for instance infinity in an asymptotically flat
situation, notice that it would require a suitable compactification)
and constructing an inertial reference frame at the reference point
(from now on we call it ``the origin''). He then specifies a second
point, not by using coordinates, but by considering a path from the
origin to the new point. To construct the path he chooses a vector
defined in the local reference frame at the origin and parallel
transports infinitesimally such reference frame along the integral
curve of the vector. At the next point another vector is chosen and so
on. For instance, one could move a certain distance along the geodesic
the $x$ direction taking the reference frame along this path, then
another distance along the $y$ direction defined with respect to the
reference frame obtained at the end of the first transport. He wishes
to describe the gravitational field in terms of these paths and
therefore without referring to a description of the space-time in
terms of coordinates defined on an open set of the space-time and
their transformations. With the information available about the paths
in this intrinsic framework one cannot say if two paths have led to
the same point just by the specification of the paths. However, the
question can be answered with a knowledge of the Riemann tensor. If
all physical measurements (e.g. all gauge invariant functions of all
fields) at the ends of the paths are the same or differ by a Lorentz
transformation we can say that they ended in the same point. It is
clear that this is not a useful way to distinguish paths in
practice. Notice that the construction is such that all along the
paths the metric is Minkowskian even though the space-time is not
necessarily flat because it results from the parallel transport of the
inertial frame $F$ given at $o$.  To have a completely invariant
description of the process, the paths are parameterized by the
invariant distance traversed (or the proper time in the case of
timelike paths).

To flesh out the above ideas, consider two paths $\pi_1$ and $\pi_2$
such that, after a portion of $\pi_2$ common to both paths (that we
shall call $\pi_3^z$) has been traversed, they differ by a small area
$\sigma_{\mu\nu}$. Given a vector $a_\mu$ in the frame at the point
where the two paths start to differ, the vector at the same point but
at the end of the closed path will differ by an amount,
\begin{equation}
  da_{\lambda}= \frac{1}{2} \sigma^{\mu\nu}
  R_{\mu\nu\lambda}{}^\sigma\left(\pi_3^z\right)
 a_\sigma. 
\end{equation}
Mandelstam denotes with $x,y,$ or $z$ the end point of the path $\pi$
in the intrinsic framework. That is, the components of the end point,
given by $x^\alpha$ are the total displacement along each of the unit
vectors of the parallel transported reference frame $e_\alpha$. The
above expression is valid in the reference frame parallel transported
to $z$ along $\pi_3^z$, we denote this by making the components of
tensors like the Riemann tensor explicitly path dependent.

The vectors defining the reference frame also get rotated and this
difference also contributes to the path dependence of the field
variables. So both vectors and the path are rotated.  If we think of
the paths as curves on space-time, the direction of the portion of
the path $\pi_2$ following $\pi_3$ will be rotated with respect to the
original portion of $\pi_1$ by an amount proportional to the Riemann
tensor at $z$.  In this framework quantities become path dependent for
two reasons: the path determines the point where the quantity is
observed and in the case of coordinate dependent quantities it also
determines the reference frame chosen to describe them.  The variation
of a vector field $A_\mu(\pi_1^x)$ in a weak gravitational field when
one moves along a path like the one described above will be given by,
\begin{equation}
\delta_z A_{\mu}(\pi_1^x)=\frac{1}{2} \sigma^{\kappa\lambda} R_{\kappa
  \lambda \mu}{}^{\nu}(\pi_3^x)A_{\nu}(\pi_1^x)-\frac{1}{2}
\sigma^{\kappa\lambda}R_{\kappa \lambda \tau}{}^{\nu}(\pi_3^z)\left(x-z\right)^{\tau} \frac{\partial A_{\mu}(\pi_1^x)}{\partial x^{\nu}}
\end{equation}
The first term is due to the rotation of the reference frame.  The
second term represents the effects of the change of the path. 
The above expression is only valid in the linearized case, it ignores
higher corrections in the curvature  and
assumes that points $x$ and $z$ are on the same flat patch
in which one can set up coordinates such that quantities like
$(x-z)^\tau$ behave as vectors and one can compute a derivative
without a non-trivial connection. 
In the
general case of a strong gravitational field there would be terms with
higher order powers in the curvature all along the path and one does
not have a closed form for the deformation at the end of $\pi_3$.  In
particular it would be very difficult to determine the displacement of the
end points under arbitrary deformations. We conclude from this
analysis that paths ending at the same physical point cannot be easily
recognizable in the intrinsic notation.  Teitelboim \cite{teitelboim}
made some progress on this issue but only for infinitesimally close
paths. Moreover, as the previous
analysis shows, the end points of two different paths like $\pi_1$ and
$\pi_2$ defined intrinsically could be the same without implying that
both paths end at the same physical point.  Another related  important obstacle
for a practical implementation of this intrinsic formalism is that the
previous analysis shows that closed loops in space-time will be very
difficult to recognize in the intrinsic notation and therefore the
groups of loops will not be of any practical use.

\section{A new intrinsic description: the group of loops in the
  gravitational case}

At the end of the previous section we have sketched some of the
obstacles faced by the Mandelstam formulation. Here we will tackle
these issues. In first place we will refine the intrinsic description
of the paths in such a way that ``trees'', that is, closed paths from
the base point $o$ equivalent to the null path that do not contribute
to holonomies, could be easily recognized.  Then we will introduce a
technique allowing to assign to each physical point, that is to each
point of the manifold $M$, intrinsically described paths that end at
that point.  These conditions will allow to apply the loop techniques
to the intrinsic description of gravitation.  In particular they will
allow to recognize closed loops in $M$ and to recognize paths ending
at the same physical point.

Let us start by a path in a manifold $M$ whose geometry is given.  We
shall assume that all the paths start at the same point $o$ of $M$. If
the space-time is asymptotically flat we shall choose $o$ at infinity
and assume diffeomorphisms and gauge transformations reduce to the
identity there.  In non asymptotically flat situations, like
cosmologies, one could pick a point in the infinite past or future
(notice that we are considering spatio-temporal paths). We will
describe paths in $M$ intrinsically in terms of a Lorentz reference
frame in $o$. Given a reference frame $F$ in $o$ a path is described
as follows: Starting from the origin we parallel transport, for an
invariant distance $ds$, the reference frame with ``velocity''
$v^\alpha(0)$ to a new point $d_1x^\alpha$ such that the displacement
is $d_1x^\alpha=v^\alpha(0)ds$. Starting at this point we proceed to
a new point moving further the reference frame with velocity
$v^\alpha(ds)$ and displacement $d_2x^\alpha=v^\alpha(ds)ds$. All the
displacements are given in terms of invariant distances and the
parallel transported reference frame. The intrinsic description of the
path $\pi^x$ may be therefore described by $x^\alpha(s)$ such that
$v^\alpha(s)=dx^\alpha/ds$ and $x^\alpha=x^\alpha(s_f)$ is the
intrinsic
total displacement associated to the end
point.  We will say that a path is reducible if it contains a portion
$x^\alpha(s)$ with $s_0<s<s_1$ such that for any point s in this
interval $v^\alpha(s)=-v^\alpha(2 s_1-s)$.  The construction is such
that portions of the path followed forward and back along the same
curve ---following a tree--- can be eliminated from the final
description of the path. This is because after following a tree one
returns to the same initial frame.  We will therefore only consider
irreducible paths under the equivalence by trees.  It will be
convenient in certain occasions to use a generic parametrization
$x^\alpha(\lambda)$ with $\lambda$ an arbitrary parameter. 
The invariant distance may be always recovered by
considering $ds=\sqrt{\eta_{\alpha\beta}dx^\alpha dx^\beta}$. At this 
stage we are not considering null paths, except perhaps as limits.

\subsection{The loop derivative}

We have already noticed that in the Mandelstam construction paths
ending at the same physical point cannot be easily recognized. They
may be identified only indirectly by noticing that all the physical
fields defined at the end of two paths ending at the same physical
points $\pi^{x_1}$ and $\pi^{x_2}$ are related by a Lorentz
transformation. Furthermore this difficulty implies that closed loops
in physical space will appear as open in intrinsic notation and that
there will be hidden relations between path dependent fields ending on
different points in intrinsic notation that extend the Eq. (4.2) to
the case of strong gravitational fields.  Without a satisfactory
solution to this problem, the approach proposed by Mandelstam cannot
be used in practice.

This difficulty can be solved as follows: given an intrinsically
described path $\pi^x$ that arrives to some physical point in $M$, we
are going to show here how to identify other intrinsic paths
$\pi^{'x'}$ that arrive at the same physical point\footnote{Notice
  that $x$ and $x'$ will be different in general, the intrinsic total
  displacements for different paths with the same end point will in
  general be different in strong gravity. We use this notation only
  for labeling points along a given path. Also notice that the
  information about the intrinsic total displacement is redundant
  because it is contained in the information that defines the path
  $\pi$, as we noted in the introduction of this section.}. This
identification will allow solving the above mentioned problems and
applying the loop calculus techniques summarized in the first
sections. Let us start by learning how to describe intrinsically
closed paths that correspond to the infinitesimal generators of the
group of loops, the loop derivatives.  The corresponding holonomies
associated with these paths determine the Lorentz transformation
connecting the reference frame $F$ given initially at $o$ with the
frame obtained at the end of the closed path.  Recall that the
infinitesimal loop added by the loop derivative using the standard
notation of section II on a differential manifold $M$ is given by
$\pi_o^x \delta \gamma \pi_x^o$ with $\delta \gamma$ obtained
traversing the curve $\delta u \delta w \overline{\delta u}
\overline{\delta w}$. However, if we describe this loop using the
intrinsic description given above in terms of displacement vectors
referred to a local system of reference parallel transported from the
origin $o$ to each point of the path after following the closed loop
$\delta \gamma$ the reference system will be rotated and, consequently, a vector at
$x$ before the rotation will be rotated by an amount $\delta
v^\rho=\delta u^\alpha \delta w^\beta R_{\alpha \beta\sigma}
{}^{\rho} v^\sigma$.  This rotation as we have discussed
implies that if one attempts going back to the origin following
$\pi_x^o$ with the same prescription given to reach $x$ in reverse
order one will end up in a different point of the space-time as shown in
figure 6.
\begin{figure} 
\includegraphics[height=6cm]{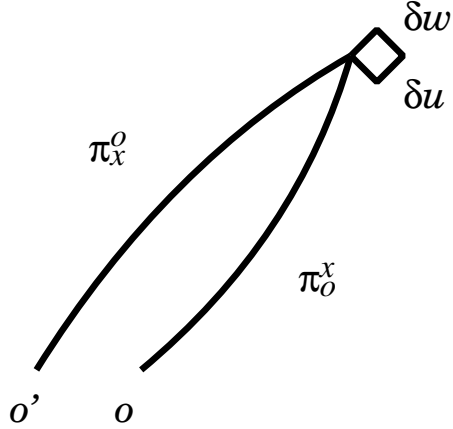}
\caption{The path described in the text: the initial and final point
  will have the same intrinsic coordinate (that is why both paths are
  labeled by $o$ and $x$) but would correspond to two
  different end points of the space-time, $o$ and $o'$.}
\label{condef} 
\end{figure} 

In order to go back to the origin along the original path in $M$, we
need to take into account the Lorentz rotation suffered by the
reference frame after following the closed path, then instead of
considering the intrinsic initial displacement $–v^\alpha(s) ds$
followed in the opposite direction we consider $(-v^\alpha(s) - \delta
u^\rho \delta w^\sigma R_{\rho \sigma} {}^{\alpha}{}_ {\beta}(\pi_0^x)
v^\beta(s)) ds$. With this prescription we are now
following the physical path $pi_o^x$ in the opposite direction,
but now, as the parallel
transported reference frame was rotated, the intrinsic displacements
needed to keep track of this rotation were rotated in the opposite
sense.  It is important to remark that when one
is back at the origin one ends up with a reference frame $F'$ rotated
with respect to the original one.  Vector components $v^\beta$ with
respect to $F$ will be related to vector components with respect to
$F'$ by a Lorentz transformation given by the holonomy,
\begin{equation}
H(\pi_o^x\circ \delta{\gamma}\circ \Lambda(\delta{\gamma})\pi_x^o)^\alpha{}_\beta=
\delta^{\alpha}{}_\beta + \delta u^\rho \delta  w^\sigma R_{\rho
  \sigma}{}^{\alpha}{}_\beta(\pi_o^x),
\end{equation}
where $\Lambda(\delta \gamma)\pi_x^o$ is the retraced rotated path
described above.


Also notice that we have followed a
closed path in space-time but the final intrinsic coordinate will
be different from the –vanishing-initial one. The intrinsic path
associated to the infinitesimal generator of the group of loops may be
represented in compact notation as $\pi \circ \delta{\gamma}\circ
\Lambda(\delta{\gamma}) \overline{\pi}$. It will be convenient in
order to keep
track of the order of infinitesimals to introduce a 
parameter $\epsilon$ with dimensions of length, much smaller than the
length associated with the curvature of space-time such that 
 $\delta u=\epsilon u$ and 
$\delta w =\epsilon w$. 

Note that the paths $\epsilon u \epsilon w \epsilon \overline{u}
\epsilon \overline{w}$ are only closed for infinitesimal loops, for
finite ones they are not closed in a generic curved space-time. In
order for it to close ---for a small, but fixed, $\epsilon$--- one has
to consider $\epsilon u \epsilon w \epsilon \overline{u}
\epsilon \overline{w^{(1)}}$ where $w^{(1)}$ is given in the appendix. The
holonomy induced by both paths coincides at order $\epsilon^2$ but
differs by terms $\left(R \epsilon^2\right)^2$ with $R$ the typical
scale of the curvature of space-time. The proposed description is
therefore correct for closed paths with finite $\epsilon$ if
$R\epsilon^2\ll 1$, which always holds for classical gravity for
sufficiently small $\epsilon$. In the quantum case one expects that $\epsilon$ cannot
be made smaller than the Planck length $\ell_{\rm Planck}$ and $R
\ell_{\rm Planck}^2$ could be of order one; for instance, in the
region of a black hole corresponding to the classical
singularity. This indicates that at those scales the notion of
curvature, and consequently the notion of point is completely lost.

In the appendix the path that must be followed to close an intrinsic
loop is constructed. The result that is convenient to keep in mind in
what follows is,
\begin{equation}
   w^{(1) \mu}=w^\mu+\frac{1}{6} R^\mu_{\alpha \rho\gamma}  w^\alpha u^\rho w^\gamma \epsilon^2+\frac{1}{3} R^\mu{}_{\alpha \rho \gamma} u^\alpha u^\rho w^\gamma \epsilon^2.
\end{equation}

It is important to point out that once one has identified closed
infinitesimal paths one has everything needed in order to describe
generic closed paths ---loops--- and in terms of them to define a
notion of point by associating them with equivalence classes of open
paths that differ by closed loops. The notion of closed path that is
proposed stops being valid when the notion of point does. This will
occur in the deep quantum regime.

Having defined intrinsic descriptions for the infinitesimal generators
of the group of loops and the associated holonomies, we can compute
the holonomies corresponding to finite deformations by considering the
product of infinitesimal generators. Notice that in order to compute
the product we need to relate each infinitesimal path to the parallel
transported reference frame the path that comes before it.  In compact
notation, for the product of two infinitesimal generators, we need to
consider the closed path,
\begin{equation}
\pi_1 \circ\delta{\gamma_1}\circ\Lambda\left(\delta \gamma_1\right) \overline{\pi_1}\circ
\Lambda\left(\delta \gamma_1\right)\pi_2\circ\Lambda\left(\delta \gamma_1\right) \delta{\gamma_2} \circ\Lambda\left(\delta \gamma_2\right)\Lambda\left(\delta \gamma_1\right) \overline{\pi_2}
\end{equation}
which corresponds to the infinitesimal holonomy $H= H_1H_2$.  Notice
that though the group of loops can be defined in an arbitrary
differential manifold (as we showed in section 2) without reference to
its geometry, the intrinsic loop description depends on the geometry.
Taking into account the way we have proceeded to compute the product
of infinitesimal generators, given two loops $\gamma_1$ and $\gamma_2$
with origin $o$ described in intrinsic notation, one can define a
product $\gamma_1\cdot \gamma_2$ given by following $\gamma_1$ and
taking into account the rotation of the reference frame at $o$, then
following $\Lambda(\gamma_1) \gamma_2$. This last object represents
the loops whose intrinsic displacements are rotated by $\Lambda$ from
the original components.  Explicitly, we have that
$\gamma_1\cdot \gamma_2=\gamma_1 \circ \Lambda(\gamma_1)\gamma_2$, and
one can easily convince oneself that intrinsic loops form a group.
The generalized Stokes' theorem allows to obtain the holonomy for an
arbitrary loop as a product of infinitesimal Lorentz transformations
associated to the infinitesimal generators.  With this definition of
the group of loops one can recognize two paths ending at the same
physical point. Two paths $\pi$ and $\pi'$ end at the same point if
there exists a loop $\gamma$ such that the open paths
$\gamma \cdot \pi=\pi'$.

\subsection{The connection derivative}

\subsubsection{A particular case}
The fact that the intrinsic
description depends on the geometry now implies that the criterion used
to recognize that two paths end in the same point does so too. 
Therefore in an eventual quantum treatment {\em the notion of point
  only acquires precise meaning when quantum
fluctuations can be neglected.} We do not
include in $\pi$ the information about the intrinsic coordinates of its
end point because these coordinates may take arbitrary values for the
same physical end point and do not add relevant information. If the
manifold is not simply connected besides the infinitesimal generators
one needs information about at least one holonomy of a loop $\gamma$ 
connecting paths
$\pi$ and $\pi'$ ending at the same physical point such that
$\gamma$ is a generator of the homotopy group.  The equivalence
class of paths that end in the same physical point may be represented
by any of the paths that end in that point.

\begin{figure}[h]
\includegraphics[height=8cm]{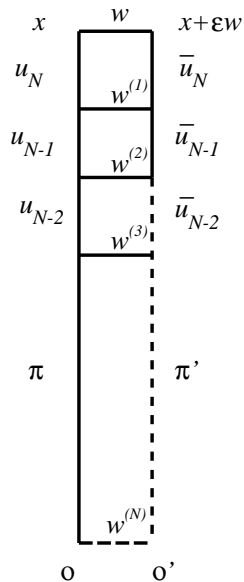}
\caption{The holonomy associated with the connection derivative. }
\label{fig1_10}
\end{figure}

We are now going to compute the holonomy associated to a
connection derivative, as in (3.5).  We will essentially reconstruct
figure 4 for a particular path using (2.19). The latter transforms 
the path $\pi_o^{o'}\circ \pi'_{o'}{}^{x+\epsilon w}$ to the path
$\pi^{x+\epsilon w}_o$ as shown in 
figure \ref{fig1_10}, where $o$ is the origin (see below for a more
precise discussion of the frames involved).  It is computed 
considering a partition $\epsilon u_1\cdots
\epsilon u_N$ of the path $\pi_o^x$ and taking the product of loop
derivatives,
\begin{equation}
  \pi_o^{x-\epsilon u_N} \circ \epsilon u_N\circ \epsilon w \circ \epsilon
  \overline{u_N}\circ \epsilon\overline{w^{(1)}} \circ 
  \epsilon w^{(1)} \circ \epsilon\overline{u_{N-1}}\circ 
  \epsilon\overline{w^{(2)}} \ldots
\end{equation}
where 
$\pi_o^{x-\epsilon u_N}$ is the portion of $\pi$ going from $o$
to $x-\epsilon u_N$. This corresponds to the transformation,
\begin{equation}
  \left(\delta_\gamma^\eta+\epsilon^2 u_N^\alpha w^\beta
    R_{\alpha\beta\gamma}{}^\eta\left(\pi_o^{x-\epsilon u_N}\right)\right)
\left(\delta_\eta^\rho
    -\tilde{u}_{N-1}^\alpha \tilde{w}^{(1)\beta} \epsilon^2
    R_{\alpha\beta\eta}{}^\rho\left(\pi_o^{x-\epsilon
      u_N-\epsilon u_{N-1}}\right)\right)\times\cdots
\end{equation} 
and taking into account that the variables with a tilde are Lorentz
transformed from the initial ones
(e.g. $\tilde{u}^\alpha=\Lambda_1{}^\alpha_\beta u^\beta$
with $\Lambda_1=\Lambda\left(\epsilon u_N\circ \epsilon w \circ \epsilon
  \overline{u_N}\circ \epsilon\overline{w^{(1)}} \right)$) we get,
\begin{eqnarray}
  &&\left(\delta_\gamma^\eta+\epsilon^2 u_N^\alpha w^\beta
    R_{\alpha\beta\gamma}{}^\eta\left(\pi_o^{x-\epsilon u_N}\right)\right)
\left[\delta_\eta^\rho+\epsilon^2\left(\delta^\nu_\sigma
    -u_N^\kappa w^\beta \epsilon^2
    R_{\kappa\beta\sigma}{}^\nu\left(\pi_o^{x-\epsilon
      u_{N}}\right)\right)\times\right.
\nonumber\\
&\times&\left. u_{N-1}^\sigma\left(\delta^\lambda_\mu -\epsilon^2 u_N^\chi w^\tau
R_{\chi\tau\mu}{}^\lambda\left(\pi_o^{x-\epsilon u_N}\right)\right)w^{(1)\mu}
R_{\nu \lambda\eta}{}^\rho\left(\pi_o^{x-\epsilon u_N-\epsilon u_{N-1}}\right)\right]
\ldots,
\end{eqnarray}
and observing that the corrections introduced by $w^{(1)}, \ldots ,w^{(N)}$ grow
with the square of the proper distance to the end point $x$ as shown in
the appendix, and keeping the result up to order linear in
$\epsilon$, we get,
\begin{equation}
  H_\gamma{}^\nu\left(\epsilon u_1\circ \ldots u_N \circ \epsilon w
    \circ \epsilon
    \overline{u_N}\circ \ldots \circ \epsilon \overline{w^{(N)}}\right) = 
\delta_\gamma^\nu +\epsilon w^\rho A_\rho{}_\gamma{}^\nu
\left(F,\pi_o^x\right),
\end{equation}
with,
\begin{eqnarray}
&&A_{\rho\gamma}{}^\nu\left(F,\pi_o^x\right) = 
\int_0^{s_f} ds \dot{y}^\alpha\left(s\right) R_{\alpha \rho
  \gamma}{}^\nu\left(\pi_o^{y\left(s\right)}\right)\nonumber\\
&&+\frac{1}{6} \int_0^{s_f} ds"\, \int_{s_f}^{s"}\, ds' \int_{s_f}^{s'}\, ds 
R_{\rho\left(\beta\alpha\right)}{}^\mu\left(\pi_o^{y\left(s\right)}\right)
\dot{y}^\beta\left(s\right)\dot{y}^\alpha\left(s'\right) 
R_{\mu\delta\gamma}{}^\nu\left(\pi_o^{y\left(s"\right)}\right) 
\dot{y}^\delta\left(s"\right)\label{5.8}
\end{eqnarray}
where the integral is along $\pi$ and 
$A_{\rho\gamma}{}^\nu$ are the Lorentz intrinsic components of
the spin connection
that depends on the path $\pi$ referred to the frame $F$. We add the
dependence on $F$ explicitly in the connection since in further usage
we will use other frames to which the specification of 
the paths are referred to. 
It is important to remark that at order
$\epsilon$ the quantities $\tilde{u}$ and $\tilde{w}$ are
equal to $u$ and $w$. 
The loop $\epsilon u_1\circ\ldots \circ \epsilon u_N \circ \epsilon w
\circ
\epsilon
\overline{u_N} \circ \ldots \circ \epsilon \overline{u^{(1)}}\circ
 \epsilon \overline{w^{(N)}}$ connects the path $\pi
\circ w$
 referred to the frame $F$
with the path $\epsilon w^{(N)} \circ \pi_{o\,'}$ referred to the frame $F'$ that
differs from $F$ by the Lorentz transformation (5.7). Both paths end at
the same physical point. 

\subsubsection{The general case}

The previously defined connection derivative is a particular example
of connections relating two ``parallel'' neighboring paths. But more generally, one can
define a connection derivative for each tangent vector in the path
manifold. If a path $\pi_o^x$ is defined 
by $u^\alpha(\lambda)=dx^\alpha(\lambda)/d\lambda$
in
the intrinsic frame parallel transported to 
the point $x^\alpha(\lambda)$, 
the tangent at the element $\pi_o^x$ of the manifold of intrinsic paths may be described by
the vector field $w^\alpha(\lambda)$ as
shown in figure 8.

\begin{figure}[h]
\includegraphics[height=6cm]{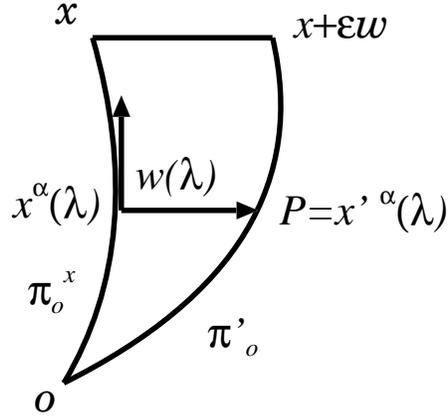}
\caption{The path defining the connection derivative. }
\label{condef2} 
\end{figure}

Let us therefore compute the holonomy associated with a generic connection
derivative, going from the path $\pi_o^x \circ w$ to the
path $\pi'_o{}^{x+\epsilon u}$ as shown in figure (\ref{condef2}), where $o$ is the
origin. Let us introduce the tangent vector at each point 
$x^\alpha(\lambda)$
of
$\pi_o^x$, given by $u^\alpha(\lambda)$. The invariant length $s$ goes
from $0$ at $o$ to $s_f$ at $x$ and
$ds=\sqrt{\eta_{\alpha\beta}u^\alpha u^\beta} d\lambda$. The path
$\pi'_o$ admits a description in terms of displacements $\epsilon
w^\alpha(\lambda)$ referred to the frame transported to the point
$x^\alpha(\lambda)$ of the path $\pi_o^x$. Different displacements $\epsilon
w^\alpha(\lambda)$ with the same final value $w^\alpha(\lambda_f)=w^\alpha$
define different connection derivatives. 

It is easy to see that \cite{teitelboim} the frame transported
up to $x^\alpha(\lambda)$ by $\pi_o^x$ and from there along $w^\alpha(\lambda)$ till
$P$ differs from the one transported along $\pi'_o$ by the
infinitesimal Lorentz transformation,
\begin{equation}
  \Lambda^\alpha{}_\beta = \delta^\alpha{}_\beta +\Omega^\alpha{}_\beta\left(\lambda\right),
  \end{equation}
with,
\begin{equation}
\Omega^\alpha{}_\beta\left(\lambda\right) = \int_o^\lambda \epsilon
R_{\gamma\delta}{}^\alpha{}_\beta\left(\lambda'\right)
u^\gamma(\lambda') w^\delta(\lambda') d \lambda'. 
\end{equation}

We can also compute $u'(\lambda)$ (the tangent to $\pi'$) in terms of $u(\lambda)$ and
$w(\lambda)$ as,
\begin{equation}
  \label{eq:3}
u'{}^\alpha=\Lambda^\alpha{}_\beta u^\beta(\lambda)+\epsilon \frac{dw^\alpha}{d\lambda},
\end{equation}
which allows to define intrinsically the path $\pi'_o$
by $u'{}^\alpha(\lambda)=dx'{}^\alpha(\lambda)/d\lambda$. The connection
derivative 
of a path dependent vector field $B^\beta(\pi)$
is given by,
\begin{equation}
  \label{eq:4}
  B^\beta\left(\pi_o^x \epsilon w(\lambda_f) \pi'{}_{x+\epsilon
      w}^o \pi\right) = \left(1+\epsilon w^\alpha(\lambda_f)
    \delta_\alpha\left(\pi_o^x\right)\right) B^\beta(\pi) =
  \Lambda^\beta_\sigma\left(\lambda_f\right) B^\sigma\left(\pi\right),
\end{equation}
and therefore,
\begin{equation}
  \label{eq:5}
  \delta_\alpha \left(\pi_o^x\right)
  B^\beta\left(\gamma\right)=A_\alpha{}^\beta{}_\sigma\left(\pi_o^x\right) B^\sigma\left(\gamma\right),
\end{equation}
with $\epsilon w^\alpha
A_\alpha{}^\beta{}_\sigma\left(\pi_o^x\right)=
\Omega^\beta{}_\sigma\left(\lambda_f\right)$.

As a consequence, choosing displacement vectors
$w^\beta(\lambda)=w^\alpha(\lambda_f) E_\alpha^\beta(\lambda)$
with $E_\alpha^\beta$ a linear transformation such that 
the evaluation of $E_\alpha^\beta$
in $\lambda_f$ is 
$E_\alpha^\beta=\delta_\alpha^\beta$ one gets,
\begin{equation}
  \label{eq:6}
A_\alpha{}^\beta{}_\sigma\left(\pi_o^x\right)=\int_o^{\lambda_f}
R_{\gamma\delta}{}^\beta{}_\sigma\left(\lambda'\right)u^\gamma\left(\lambda'\right)
E_\alpha^\delta\left(\lambda'\right) d\lambda'=\int_o^{s_f}
R_{\gamma\delta}{}^\beta{}_\sigma\left(y\right) 
E_\alpha^\delta\left(y\right) d {x}^\gamma,
\end{equation}
with $dx^\alpha= u^\alpha\left(\lambda\right)d\lambda$ and 
the integral is along $\pi_o^x$ referred to the frame $F$. Notice that the
connection derivative is not unique and would require to include the
information
about $E^\beta_\alpha(\lambda)$ 
for $0\le \lambda \le \lambda_f$ with the fixed
boundary condition
$E_\alpha^\beta(\lambda_f)=\delta_\alpha^\beta$. The complete notation
would therefore be $A_\alpha{}^\beta{}_\sigma\left(F,
  \pi_o^x,[E_\alpha]\right)$, where $[E_\alpha]$ defines the tangent
vector basis to the path $\pi_o^x$. 

Notice that in the definition of the connection derivative introduced
in section II.C there was an assignment of paths to the points of the
manifold. A different assignment corresponds to a gauge change. Here
that role is being played by the matrices $E_\alpha^\beta$.

\subsection{Finite deformations}

\subsubsection{A finite loop based on ``parallel'' connections}

We are now in the position to compute the holonomy associated to a
closed finite path that extends the path ordered exponentials (2.18)
and
(3.9) to the gravitational case. We first analyze for simplicity a
finite loop generated by ``parallel'' connections. 
This relationship allows to obtain $H^\alpha{}_\beta$ as 
a path ordered exponential. 
The construction that follows can be
done with the connection (5.7) or the ones stemming from 
the connection (5.14) associated to figure 8. 
\begin{figure}[h]
\includegraphics[height=10cm]{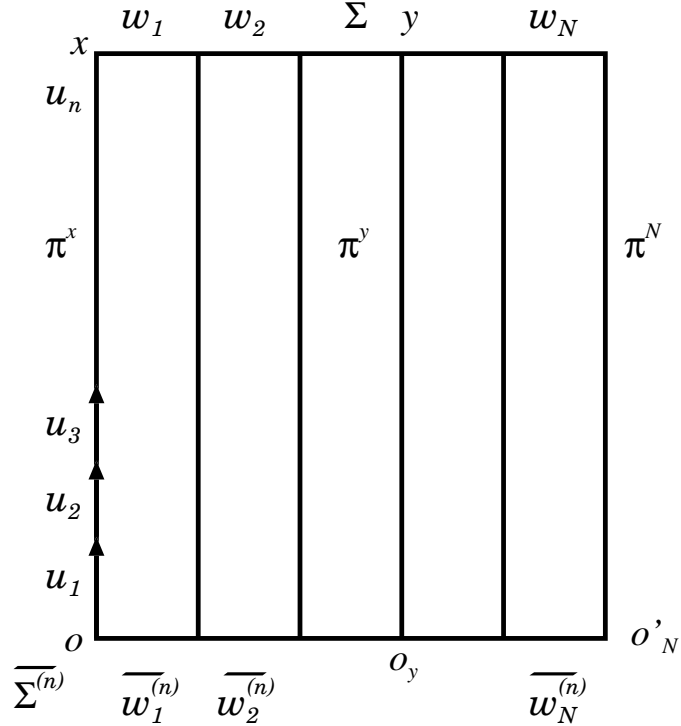}
\caption{The path $\gamma=\pi^x \circ\Sigma \circ\overline{\pi^{N}}\circ \overline{\Sigma^{(n)}}$ used
  in the construction of the holonomy associated to closed finite path.}
\label{fig9}
\end{figure}

To obtain a closed path in intrinsic gravity is non-trivial but
crucial for identifying physical points in the manifold. The idea is to
construct them by composition of paths associated to connections like
those in figures 7 and 8. We wish to define the path of figure 9 in
intrinsic notation as a loop referred to the frame $F$, appropriately parallel transported.
Omitting the $\epsilon$'s it is given by $\gamma = \pi^x \circ
w_1\circ\ldots \circ w_N\circ
\overline{\pi}\circ  \overline{\Sigma}^{(n)}_{o o'_N}$. The idea is to
obtain it as a product of
infinitesimal deformations that we organize in brackets, as shown in
figure 9, 
\begin{eqnarray}
  \gamma &=& \left( \pi^x\circ w_1\circ \overline{\pi}^{y_1}\circ\overline{w_1}^{(n)}\right)\vert_F
\left.\left( w_1^{(n)} \circ {\pi}^{y_1} \circ w_2\circ
    \overline{\pi^{y_2}}\circ\overline{w_2}^{(n)}\Lambda_2
    \,\overline{w_1}^{(n)}\right)\right\vert_{F_1} \nonumber\\
&&\times \left.\left( \Lambda_2 w_1^{(n)} \circ w_2^{(n)} \circ \pi^{y_2} \circ w_3
  \circ \overline{\pi^{y_3}}\circ
  w_3^{(n)} \circ \Lambda_3 \overline{w_2}^{(n)} \circ\Lambda_3 \Lambda_2
  \overline{w_1}^{(n)}\right)\right\vert_{F_2} \nonumber\\
&&\times \left.\cdots \left(\Sigma_{oo_{p}} \circ\pi^{y_p}  \circ w_{p+1} \circ
  \overline{\pi^{y_{p+1}}}  \circ \overline{w_{p+1}}^{(n)} \circ
  \overline{\Sigma_{o_{p+1}o}}\right)\right\vert_{F_p}
\cdots
\end{eqnarray}
with $\Sigma_{oo_p} =\Lambda_p\Lambda_{p-1} \cdots \Lambda_2 w_1^{(n)}  \circ
\Lambda_p \ldots 
\Lambda_3 w_2^{(n)}  \circ \cdots \Lambda_p w_{p-1}^{(n)}  \circ w_p^{(n)}$ and where
the subscript $F_p$ means the frame
rotated by $\Lambda_p \cdots \Lambda_1$ of $F$ and $\Lambda_p$ the
infinitesimal Lorentz transformation induced by the closed path
$\pi_{p-1}\circ w_p \circ \overline{\pi_p} \circ \overline{w_p}^{(n)}$
(notice the change in notation for $\Lambda$'s). 

The equation for $\gamma$ leads to an expression very similar to (3.9)
for the holonomy, 
\begin{eqnarray}
  H^\alpha{}_\beta &=& \left(\delta^\alpha_{\beta_1} +\epsilon
    w_1^\rho A_\rho{}^\alpha{}_{\beta_1}\left(F, \pi\right)\right)
\left(\delta^{\beta_1}_{\beta_2} +\epsilon
    w_2^\rho A_\rho{}^{\beta_1}{}_{\beta_2}\left(F_1,\Sigma_{oo_1}\pi_1\right)\right)
\cdots \nonumber\\
&&\times
\left(\delta^{\beta_p}_{\beta_{p+1}} +\epsilon
    w_{p+1}^\rho A_\rho{}^{\beta_p}{}_{\beta_{p+1}}\left(F_p,\Sigma_{oo_p}\pi_p\right)\right)\cdots
\end{eqnarray}
that is,
\begin{equation}
  H(\gamma) = {\rm P}\exp\left(i \int_\Sigma dy^\alpha A_\alpha\left(F_y,\Sigma_{oo_y}\pi_{o_y}^y\right)\right).
\end{equation}

We therefore recover the intrinsic version of the non-Abelian Stokes'
theorem.

\subsubsection{A finite loop based on general connections}

We now proceed to construct a finite loop based on general
connections. 
The idea is to obtain $\gamma=\pi_o^x \Sigma
\overline{\pi}^N$,
as shown in figure 10, as a product of infinitesimal deformations that we organize in
brackets,
\begin{equation}
  \label{eq:19}
  \gamma=
\left(\pi_o^x \epsilon w_1 \overline{\pi}_1^{y_1}\right)
\left(\pi_1^{y_1} \epsilon w_2 \overline{\pi}_2^{y_2}\right)\ldots
\left(\pi_p^{y_p} \epsilon w_p \overline{\pi}_{p+1}^{y_{p+1}}\right)\ldots
\end{equation}

\begin{figure}[h]
\includegraphics[height=8cm]{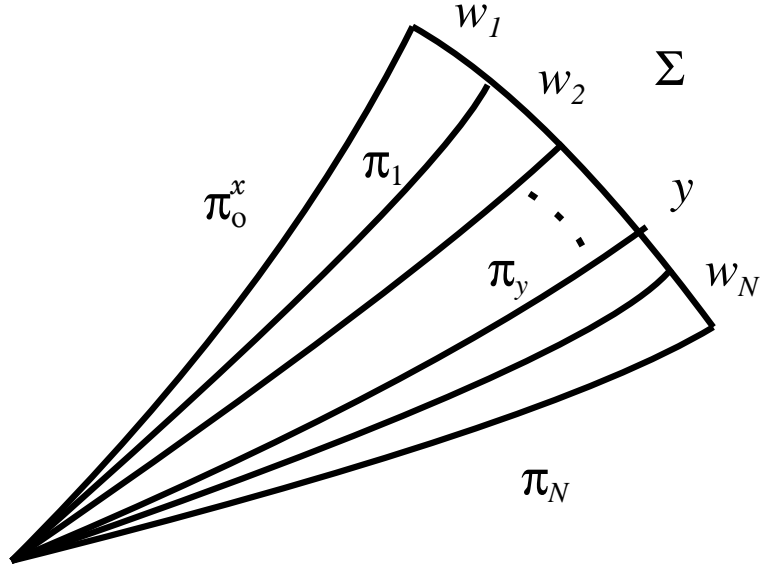}
\caption{The path $\gamma=\pi^x \circ\Sigma \circ\overline{\pi^{N}}$ used
  in the construction of the holonomy associated to closed finite path.}
\label{fig9n}
\end{figure} 

Where $w^\alpha_i=w_i^\alpha(\lambda_f)$ and $\overline{\pi}_i^{y_i}$ is the path defined by the tangent $u^\alpha_i(\lambda)$
referred to the frame parallel transported along
$\pi_o^x{}\circ \epsilon w_i$. We therefore repeat the calculations in
equations (5.9) and subsequent ones. We have that,
\begin{equation}
  \label{eq:7}
  u^\alpha_1(\lambda) = \left(\delta^\alpha{}_\beta
    +\Omega_1{}^\alpha{}_\beta(\lambda)\right)u^\beta(\lambda) +
  \epsilon \frac{dw_1^\alpha(\lambda)}{d\lambda},
\end{equation}
and 
\begin{equation}
  \label{eq:8}
  \Omega_1{}^\alpha{}_\beta(\lambda) = \epsilon \int_0^\lambda 
  R_{\gamma\delta}{}^\alpha{}_\beta(\lambda')u^\gamma(\lambda')
  w_1^\delta(\lambda') d \lambda',
\end{equation}
where $u^\gamma(\lambda)$ is the tangent vector to the path
$\pi_o^x$. 
Analogously, $\pi_p^{y_p}$ is the path given by the tangent
  vector $u^\alpha_p$ given by,
  \begin{equation}
    \label{eq:9}
    u_p^\alpha(\lambda) = \left(\delta^\alpha{}_\beta
      +\Omega_p{}^\alpha{}_\beta(\lambda)\right)
    u_{p-1}^\beta(\lambda) + \epsilon \frac{d w_{p}^\alpha}{d\lambda},
  \end{equation}
with,
\begin{equation}
  \label{eq:10}
\Omega_p{}^\alpha{}_\beta(\lambda) =\epsilon \int_0^\lambda 
R_{\gamma\delta}{}^\alpha{}_\beta(\lambda') u^\gamma_{p-1}(\lambda')
w_{p}^\delta(\lambda') d\lambda'.
\end{equation}

These relations may be written as follows, parametrizing the $u's$ in
such a way that $u^\alpha(\lambda, \mu=k\epsilon)=u^\alpha_k(\lambda)$,
and $w^\alpha(\lambda, \mu=k \epsilon)=w_k^\alpha(\lambda)$
respectively,
with $\epsilon=d\mu$, 
\begin{equation}
  \label{eq:11}
  u^\alpha(\lambda,\mu)-u^\alpha(\lambda,
  \mu-d\mu)=\Omega^\alpha{}_\beta(\lambda,\mu)
u^\beta(\lambda,\mu) d\mu+\frac{dw^\alpha(\lambda,\mu)}{d\lambda} d\mu,
\end{equation}
and,
\begin{equation}
  \label{eq:12}
\Omega^\alpha{}_\beta(\lambda,\mu) = \int_0^\lambda
R_{\gamma\sigma}{}^\alpha{}_\beta(\lambda',\mu) u^\gamma(\lambda',\mu)
 w^\sigma(\lambda',\mu) d\lambda',
\end{equation}
and therefore,
\begin{equation}
  \label{eq:13}
  \frac{du^\alpha(\lambda,\mu)}{d\mu}=\Omega^\alpha{}_\beta(\lambda,\mu)
  u^\beta(\lambda,\mu) + \frac{d w^\alpha(\lambda,\mu)}{d\lambda}.
\end{equation}

If one can solve the above equations one gets an expression for the
finite deformation. 
The expression for $\gamma$ leads to,
\begin{equation}
  \label{eq:17}
  H^\alpha{}_\beta = \left(\delta^\alpha{}_{\beta_1} +\delta w_1^\rho
    A_\rho{}^\alpha{}_{\beta_1}(\pi_o^x) \right) 
\left(   \delta^{\beta_1}{}_{\beta_2} + \delta w_2^\rho   A_\rho{}^{\beta_1}{}_{\beta_2}(\pi_1^{y_1})\right)\ldots
    \left(\delta^{\beta_p}{}_\beta + \delta w_p^\rho 
A_\rho{}^{\beta_p}{}_\beta(\pi_{p-1}^{y_{p-1}})\right)\ldots
\end{equation}
that is,
\begin{equation}
  \label{eq:18}
  H(\gamma) ={\rm P}\exp\left(i \int_\Sigma dy^\alpha A_\alpha(\pi^y)\right),
\end{equation}
with the connection given by (5.14). We 
see that in this case the intrinsic version of the non-Abelian Stokes'
theorem takes the standard form. 
The loop gamma connects the path $\pi_o^x \circ \Sigma$ with $\pi_N$, and noticing that $w_2$ is referred to the frame transported along $\pi_1$, etc.,
we get $\Sigma =  \delta w_1 \circ \Lambda_1 \delta w_2 \circ \ldots  \circ \Lambda_1
\Lambda_2 \ldots \Lambda_{N-1}\delta w_N$ with 
$\Lambda_p{}^\alpha_\beta=\delta^\alpha_\beta+\Omega_p{}^\alpha_\beta$
and $\Omega_p$ given by equation (\ref{eq:10}) . 

To compute explicitly the connection for the path $\pi^y$ one needs to
solve for $u^\alpha(\lambda,\mu)$, which requires the solution of
(\ref{eq:13}). This might be solved in closed form for particular
geometries. One can proceed to solve it iteratively for weak fields.
Let us denote by $u_{(0)}, u_{(1)}, u_{(2)}$ the
order of iteration computed, we have,
\begin{eqnarray}
  \label{eq:14}
u^\alpha_{(0)}(\lambda,\mu)&=& u^\alpha(\lambda,0),\\
  u_{(1)}^\alpha(\lambda,\mu) &=& u^\alpha(\lambda,0)+ \int_0^\mu \frac{d
    w^\alpha(\lambda,\mu')}{d\lambda} d\mu',
\end{eqnarray}
with,
\begin{eqnarray}
  \label{eq:15}
  \frac{d u^\alpha_{(2)}}{d\mu} &=& \Omega_{(1)}{}^\alpha{}_\beta
  u^\beta_{(1)} +\frac{dw^\alpha(\lambda,\mu)}{d\lambda}\nonumber\\
&=& \int_0^\lambda R_{\gamma\delta}{}^\alpha{}_\beta(\lambda',\mu)
    u_{(1)}^\gamma(\lambda',\mu) w^\delta(\lambda', \mu)
d\lambda'    u^\beta_{(1)}(\lambda,\mu) +\frac{d w^\alpha(\lambda,\mu)}{d\lambda},
\end{eqnarray}
and  
with $R_{\gamma\delta}{}^{\alpha \beta}(\lambda',
\mu)=R_{\gamma\delta}{}^{\alpha \beta}(\pi_{(1)}(\mu))$, with
$\pi_{(1)}(\mu)$ defined by  $x^\alpha_{(1)}(\lambda,\mu)$ such that
$\partial_\lambda x^\alpha_{(1 )}(\lambda,\mu)
=u^\alpha_{(1)}(\lambda,\mu)$ and 
\begin{eqnarray}
  \label{eq:16}
  u^\alpha_{(2)}(\lambda,\mu) &=& 
\int_0^\mu d\mu' 
\left\{
\int_0^\lambda  d\lambda' 
R_{\gamma\sigma}{}^\alpha{}_\beta\left(x_{(1)}(\lambda',\mu')\right)
u_{(1)}^\gamma(\lambda',\mu')
  w^\sigma(\lambda', \mu') u_{(1)}^\beta(\lambda,\mu')
\right\}\nonumber\\
 &&+ u_{(1)}^\alpha(\lambda,\mu)
\end{eqnarray}
and by iteration we determine $u^\alpha(\lambda,\mu)$ for sufficiently
weak fields.

\section{Path dependent fields}

Let us consider fields with tensor,
spinor or internal components. One can start by giving the fields for
arbitrary paths at each point $\phi^{(A, I)}(\pi)$ where the index $A$
represents the Lorentz tensor or spinor components and $I$ the internal
components. The indices refer to the components in the frame parallel
transported along the path. Having recognized the closed loops $\gamma$,
the fields transform under changes of the reference path by
representations of the group of loops. For instance for a vector field
with internal group $SU(N)$ in some representation,
\begin{equation}
A^\alpha{}_I(\pi') = H(\gamma)^\alpha{}_\beta H\left(\gamma\right)
_I{}^J A^\beta{}_J(\pi),
\end{equation}
if $\pi'=\gamma\circ\Lambda(\gamma)\pi=\gamma\cdot \pi$,
which guarantees that $\pi'$
and $\pi$ end at the same point on $M$. $H(\gamma)^\alpha{}_\beta$ is a
holonomy associated with the Lorentz group and $H(\gamma)_I{}^J$ a
holonomy associated with the internal group. The path-dependent fields like
$A^\beta{}_J$ depend on the paths $\pi$ referred to the frame $F$ chosen
as a reference at $o$.  Analogous relations hold for any matter
field and should be compared with the corresponding relation in
Mandelstam notation (4.2) that cannot even be written explicitly in the
case of strong fields.

The notion of covariant derivative of path dependent fields can be
introduced using the Mandelstam derivative. Its
meaning for gauge theories was analyzed in sections II y III, defined
by $(1+\epsilon u^\beta D_\beta) A^\alpha{}_I(\pi^z)=
A^\alpha{}_I(\pi_E^{z+\epsilon u})$. 
Where $\pi_E^{z+\epsilon u}$ is the path extended in the direction $u$ 
whose components are given with respect to the frame at the end point $z$. 
It compares the field parallel
transported from $z+\epsilon u$ to $z$ with the field at $z$ and therefore
gives us the component of the space time covariant derivative with
respect of the intrinsic basis  parallel transported along $\pi$.
$\pi_E$ is the extended path shown in figure (3) but now the extension
is given in terms of the intrinsic components of $u$ in the frame
parallel transported up to z. 

\subsection{Symmetries of the path dependent Riemann tensor}

As we mentioned in section II one can derive a Bianchi identity by
considering a tree that follows the edges of a cube and noticing that
``the boundary of a boundary vanishes''. If this construction is done
at the end point of $\pi$ one gets
\begin{equation}
\left([D_\beta[D_\gamma,D_\delta]]
+[D_\gamma[D_\delta,D_\beta]]
+[D_\delta[D_\beta,D_\gamma]]
\right)A_\alpha(\pi)=D_{[\beta}R_{\gamma \delta]
\alpha}{}^\epsilon\left(\pi\right) A_\epsilon\left(\pi\right)=0 \label{6.2}
\end{equation}
which implies that the path dependent Riemann tensor satisfies the Bianchi
identity. In the intrinsic formalism we are developing, a scalar
satisfies $\phi(\pi)=\phi(\pi')$ if $\pi'=\gamma \cdot \pi$ and,
applying the same construction with a scalar we get,
\begin{equation}
\left([[D_\alpha,D_\beta]D_\gamma]
+[[D_\beta,D_\gamma],D_\alpha]
+[[D_\gamma,D_\alpha],D_\beta]
\right)
\phi(\pi)=R_{[\alpha\beta\gamma]}{}^\delta D_\delta \phi(\pi)=0.\label{6.3}
\end{equation}
Since by construction the Riemann tensor is antisymmetric in the first
two and the last two indices, the above identities imply the remaining
algebraic identities of Riemann's tensor are all satisfied.

\begin{figure}[h]
\includegraphics[height=5cm]{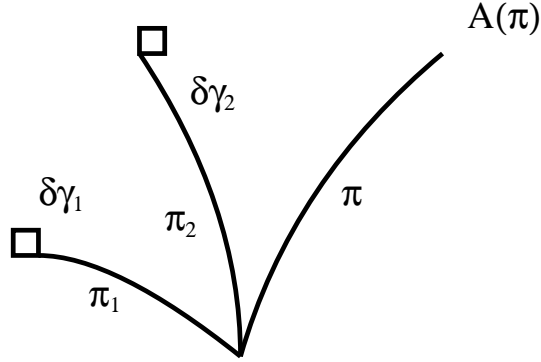}
\caption{The path used to show the Lorentz transformation of the
  Riemann tensor.}
\label{riemannlorentz}
\end{figure}

In what follows, as an application of the techniques developed up to
now, we will show that the Riemann tensor has the expected tensorial
transformation under changes of path. So we consider a one form along
a path with a small closed loop.  And then along a path with two small
loops.  The first term will give rise to a rotation of the form given
by the Riemann tensor evaluated in the path $\pi_2$ as per (5.1). The
second deformation will change the frame of the Riemann tensor, which
will therefore be Lorentz transformed by going from the path $\pi_2$
to $\delta\gamma_1\cdot \pi_2$. The paths are shown in figure
(\ref{riemannlorentz}). Let us start by computing,
\begin{equation}
  A_\alpha\left(\delta  \gamma_1 \cdot \delta  \gamma_2 \cdot \overline{\delta \gamma_1}
    \cdot \pi\right)- A_\alpha\left(\delta \gamma_2 \cdot \pi\right)
\end{equation}
where $A_\alpha$ is a path dependent intrinsic description of a one
form, $\delta\gamma_1 =\left(\pi_1 \circ \delta u_1 \circ \delta
w_1\circ \overline{\delta u_1} \circ \overline{\delta
  w_1}\circ\Lambda\left(\delta \gamma_{1x}\right)
\overline{\pi_1}\right)_F$, with 
$\delta \gamma_2$ similarly defined for $\pi_2$. Notice that for
brevity we have
slightly changed the notation in that $\delta \gamma_i$ include the
path $\pi_i$ now. We also have that $\overline{\delta
  \gamma_1} = \left(\pi_1 \circ \delta w_1 \circ \delta u_1 \circ \overline{\delta
  w_1}\circ \overline{\delta u_1}\circ 
\Lambda\left(\overline{\delta
    \gamma_{1x}}\right) \overline{\pi_1}\right)_{F_1}$ where $F_1$ is the frame rotated with
$\Lambda$ of $F$. Therefore the variation of the Riemann tensor
 under a change of path is given by,
\begin{equation}
  \sigma_2^{\eta\rho} \delta R_{\eta \rho \alpha}{}^{\beta}\left(\pi_2\right)A_\beta\left(\pi\right)=\left[\left(\delta_\alpha{}^\beta+\left(\tilde{\sigma}\right)_2^{ \eta \rho}
      R_{\eta\rho\alpha}{}^\beta\left(\delta\gamma_1\cdot
        \pi_2\right)\right)
    -
    \left(\delta_\alpha{}^\beta+\sigma_2^{\eta \rho}
      R_{\eta\rho\alpha}{}^\beta\left(\pi_2\right)\right)\right]A_\beta
  \left(\pi\right)
  ,\label{7.5}
\end{equation}
where 
\begin{equation}
\sigma^{\eta\rho}_i = \frac{1}{2} \epsilon^2 \left(
\delta u^\eta_i \delta w^\rho_i-\delta u^\rho_i \delta w^\eta_i\right)
\end{equation}
and the components of $\tilde{\sigma}^{\eta\rho}_2$ are rotated with 
$\Lambda(\delta \gamma_1)$. 
In order to compute $\delta R_{\eta \rho
  \alpha}{}^\beta\left(\pi_2\right)$, that represents the variation of $R$
under the deformation $\pi_2\to \delta \gamma_1\cdot \pi_2$, we note that
(\ref{7.5}) can be rewritten as,
\begin{eqnarray}
  &&\left[\left(\delta_\alpha{}^\lambda
      +\sigma_1^{\mu\nu}R_{\mu\nu\alpha}{}^\lambda\left(\pi_1\right)\right)\left(\left(\tilde{\sigma}\right)_2^{\delta\rho}
      R_{\delta\rho \lambda}{}^\gamma\left(\pi_2\right)
\right)
\left(\delta_\gamma{}^\beta-\sigma_1^{\mu'\nu'}
        R_{\mu'\nu' \gamma}{}^\beta\left(\pi_1\right)\right)\right.\nonumber\\
&&\left.
-
\sigma_2^{\delta\rho}
      R_{\delta\rho
        \alpha}{}^\beta\left(\pi_2\right)
\right]A_\beta\left(\pi\right)=
\sigma_2^{\eta\rho} \delta R_{\eta \rho \alpha}{}^{\beta}\left(\pi_2\right)A_\beta\left(\pi\right),
\end{eqnarray}
and taking into account that 
\begin{equation}
\left(\tilde{\sigma}\right)_2^{\rho\sigma} =
\sigma_2^{\rho\sigma}
+\sigma_1^{\mu\nu}R_{\mu\nu\epsilon}{}^\rho\left(\pi_1\right)\sigma_2^{\epsilon\sigma}
+\sigma_1^{\mu\nu} R_{\mu\nu\epsilon}{}^\sigma\left(\pi_1\right)
\sigma_2^{\rho\epsilon}, 
\end{equation}
we see that (\ref{7.5}) can be rewritten as,
\begin{eqnarray}
\delta R_{\eta \rho \alpha}{}^{\beta}\left(\pi_2\right)&=&  \left[
\omega_\alpha{}^\lambda\left(\pi_1\right)R_{\eta\rho\lambda}{}^\beta\left(\pi_2\right) 
-\omega_\lambda{}^\beta\left(\pi_1\right)R_{\eta\rho\alpha}{}^\lambda\left(\pi_2\right)
\right.\nonumber\\
&&\left.+\omega_\eta{}^\gamma\left(\pi_1\right)R_{\gamma\rho\alpha}{}^\beta\left(\pi_2\right)
+\omega_\rho{}^\gamma\left(\pi_1\right)R_{\eta\gamma\alpha}{}^\beta\left(\pi_2\right) \right]
\end{eqnarray}
with
$\omega_\alpha{}^\lambda\left(\pi_1\right)=\sigma_1^{\mu\nu}R_{\mu\nu\alpha}{}^\lambda\left(\pi_1\right)$
and $R$ undergoes a Lorentz transformation under a change of paths. 

\subsection{Equations of motion} 

To illustrate how one would write path dependent equations of motion
let us consider a gravitating scalar
field, 
\begin{eqnarray}
  \left( \eta^{\alpha\beta} D_\alpha D_\beta -m^2 \right)
  \phi\left(\pi\right)&=&0\\
R_{\alpha \lambda \beta}{}^\lambda\left(\pi\right)-\frac{1}{2}
\eta_{\alpha\beta} \eta^{\gamma\rho} R_{\gamma
  \lambda\rho}{}^\lambda\left(\pi\right) &=& \kappa
T_{\alpha\beta}\left(\pi\right), 
\end{eqnarray}
with 
\begin{equation}
  T_{\alpha\beta}\left(\pi\right) =D_\alpha \phi\left(\pi\right)
  D_\beta \phi\left(\pi\right)-\frac{1}{2} \eta_{\alpha\beta}
  \eta^{\mu\nu} D_\mu \phi\left(\pi\right)
D_\nu \phi\left(\pi\right)-m^2 \phi^2\left(\pi\right) \eta_{\alpha\beta},
\end{equation}
and $\kappa= 8\pi G$. Notice that all tensor components are Lorentzian
components in the local frame, therefore the metric is the Minkowski
one. Recall that in the intrinsic description the physical points are
associated with classes of paths that differ by loops. Although scalar
fields are only point dependent and $\phi(\pi)=\phi(\gamma\cdot \pi)$,
the information about points is given in terms of a path. The
intrinsic description of the paths ensures that $\phi(\pi)$ is a
diffeomorphism invariant physical observable. The discussion of the
next section will allow to reproduce the ordinary equations from the
path dependent ones.

\section{Recovering the standard coordinate dependent description}
\subsection{Going from the intrinsic to coordinate description}

We have shown that the intrinsic description allows to recognize when
open paths lead to the same point. Let us consider an assignment of
reference paths that define normal coordinates at each point of a
region $U$ sufficiently small around a point $P$ to which we have
arrived following a geodesic that starts at $o$. That is $P$ is
intrinsically defined following a geodesic starting at $o$ given by 
by $z^\alpha(s) = s u^\alpha$, where $u^\alpha$ is
a vector in the frame $F$. The point $P$ corresponds to $s=s_P$.  A point $Q$ of $U$
is given by $z^\alpha(Q) = s_P u^\alpha + s_Q v^\alpha$ with
$v^\alpha$ the vector components relative to 
the frame parallel transported to $P$ of the
tangent  at $P$ of the 
geodesic that joins $Q$ with $P$. The
construction is possible locally since we assume that there exists a
unique geodesic at $U$ from $P$ to $Q$. The quantities
$x^a(Q)\equiv z^a(Q)-s_P u^a$
define a chart that maps the points of $U$ to a region of $R^4$ that
are Riemann normal coordinates with origin at $P$ (we denote Riemann
coordinates with Latin letters). It is
possible to define charts $\overline{x}^a(Q)$ diffeomorphic to
$x$. The intrinsic construction allows to associate to each $Q$, in
addition to its coordinates $x^a(Q)$ the coordinates of the local
frame transported from $o$ to that point
$e_\alpha^a\left(\pi_R^Q\right)$ with $\pi_R^Q$ the above mentioned
path going from $o$ to $P$ and from there to $Q$.

The frames transform under changes of path $\pi_R^Q\to
\pi'{}^Q$ (keeping the original coordinates defined by $\pi_R$) as,
\begin{equation}
  e_\beta^a\left(\pi'{}^Q\right) = H_\beta{}^\alpha\left(\gamma \right) e_\alpha^a\left(\pi_R^Q\right) = 
 H_\beta{}^\alpha\left(\gamma
    \right) e_\alpha^a\left(x(Q)\right), \label{7.1}
\end{equation}
with $H_\beta^\alpha$ the Lorentz transformation associated with the
holonomy along the closed loop $\gamma$ is such that
$\pi'{}^Q=\gamma\cdot \pi^Q_R$. Recall that the index $\alpha$ corresponds to a
frame index and the index $a$ is a coordinate index.  Under
diffeomorphisms $x^a\to \overline{x}^a(x)$, we have that,
\begin{equation}
  e_\alpha^b\left(x'\right)= \frac{\partial x'{}^b}{\partial x^a}
  e_\alpha^a\left(x\right). 
\end{equation}
The metric in this system of coordinates can be specified as usual in terms of tetrads,
\begin{equation}
  g^{ab}\left(x\right) = \eta^{\alpha\beta}
  e_\alpha^a\left(\pi_R^x\right) e_\beta^b\left(\pi_R^x\right),
\end{equation}
and is independent of the reference path since the holonomies are
Lorentz transformations. Since the tetrads are
obtained by parallel transport from the origin, and taking into account
the definition of the Mandelstam derivative, the intrinsic construction
implies immediately that 
\begin{equation}
  D_\alpha e_\beta^b\left(\pi_R^x\right)=0. \label{triangulo} 
\end{equation}

Defining,
\begin{equation}
  \nabla_a e_\beta^b\left(x\right) \equiv e_a^\alpha D_\alpha
  e_\beta^b\left(\pi_R^x\right),
\end{equation}
we have that $\nabla_a e_\beta^b\left(x\right)=0$ and we recover the
usual covariant derivative since it compares the tetrad at $x+dx$ with
the parallel transported one at that point. As a consequence $\nabla_a
g^{bc}\left(x\right)=0$ and the connection is metric compatible. 

To show that the torsion is zero we consider a scalar field
$\phi\left(x\right)=\phi\left(\pi_R^x\right)=\phi\left(\pi_R^Q\right)$. We
have that $\phi\left(\pi_R^Q\right)=\phi\left(\pi'{}^Q\right)$ 
for any path $\pi$ arriving at $Q$,
and taking into account the intrinsic version of (\ref{ricciid}), we have that,
\begin{equation}
  D_{[a}D_{b]}\phi\left(\pi\right)=\frac{1}{2}
  \Delta_{\alpha\beta}\left(\pi\right) \phi(\pi)=0
\end{equation}
since $\phi$ is really path independent, 
and therefore the connection is therefore torsion free. 

By construction, since the point $P$ is the origin of the normal
coordinates we are using, we have at $P$ that
$e_\alpha^a\left(\pi_R^P\right)=e_\alpha^a\left(P\right)=\delta_\alpha^a$
and for $Q$, using well known results for normal coordinates we have
that,
\begin{equation}
  e_\alpha^a\left(x_Q\right)=e_\alpha^a\left(\pi_R^Q\right)=\delta_\alpha^a
  + \frac{1}{3} R^a_{b \alpha c}\left(\pi_R^P\right)x^bx^c +O\left(s_Q^3\right),
\end{equation}
recalling that at second order in Riemann coordinates the Riemann
tensor is evaluated at the origin 
$P$ where intrinsic and Riemann components coincide.

Although the Riemann tensor identities follow from the intrinsic ones
given in VIa from the metricity and torsion freedom of the connection,
it is immediate to obtain the identities in terms of coordinates from
the intrinsic ones taking into account (\ref{triangulo}), and the
discussion presented in section VIa, and recalling that at P the tetrad components in Riemann coordinates reduce to the identity.

\subsection{Relating intrinsic and coordinate descriptions of paths and local frames}

We would like to relate the paths described in coordinate systems with
intrinsic paths and identify the local frames at an arbitrary point of
the path in terms of the geometric or intrinsic descriptions of the
paths. Let $\gamma^a(\lambda)$ be a curve in an arbitrary coordinate
system such that $\gamma^a(0)=x_o^a$, 
the coordinates of $o$, and
$\gamma(1)=x^a$. We want to determine
$e_\alpha{}^a(\lambda=1)=e_\alpha{}^a(\gamma(\lambda=1))$ and in
general $e_\alpha{}^a(\lambda)=e_\alpha{}^a(\gamma(\lambda))$ and from
them the
intrinsic components of $\gamma^a(\lambda)$, let us call them
$y^a(\lambda)$. 

Using that,
\begin{equation}
d\lambda\dot{\gamma}^a\nabla_a e_\alpha{}^b = d\lambda \dot{\gamma}^a
\left(\partial_a +\Gamma_{ad}{}^b\right)e_\alpha^d=0,
 \end{equation}
it follows that,
\begin{equation}
  e_\alpha{}^c(\lambda+d\lambda) = \left(\delta^c_d-d\gamma^a \Gamma_{ad}{}^c\left(\gamma(\lambda)\right)\right)e_\alpha{}^d,  
\end{equation}
which can be integrated along the path to give,
\begin{equation}
  e_\alpha{}^c(\lambda)={\rm P}\left(\exp\left(-\int_0^\lambda
      d\lambda'\dot{\gamma}^a(\lambda') \Gamma_a\right)\right)_d^{\,\,\,c} e_\alpha{}^d(0),  
\end{equation}
and for $e_\alpha{}^d(0)=\delta_\alpha^d$ one gets the explicit form of the parallel transported local frame along gamma,
\begin{equation}\label{7.11}
  e_\alpha{}^c(\lambda)={\rm P}\left(\exp\left(-\int_0^\lambda
      d\lambda'\dot{\gamma}^a(\lambda') \Gamma_a\right)\right)_\alpha^{\,\,\,c},
\end{equation}
and the intrinsic coordinates are
\begin{eqnarray}
  \frac{d y_\alpha}{d\lambda}&=&\dot{\gamma}_c(\lambda)
                                 e_\alpha{}^c(\lambda),\\
y^\alpha(\lambda)&=& \int_0^\lambda \dot{\gamma}^c(\lambda')
                     e^\alpha{}_c(\lambda') d\lambda'.
\end{eqnarray}

Knowing the geometry, the metric in $M$ and its associated connection
allows to determine through (\ref{7.11}) the intrinsic coordinates
associated to any given curve $\gamma$.

The inverse correspondence allows to associate to each  path
$\pi$, described intrinsically by $y^\alpha(\lambda)$ and each system of
coordinates, the components of the frame parallel transported along
$\pi$ and the curve in coordinates $\gamma^a(\lambda)$ that
corresponds to the intrinsic path $y^\alpha(\lambda)$, 
\begin{eqnarray}
  e_\alpha{}^a(\lambda)&\equiv& e_\alpha{}^a\left(\left[y^\beta\right],\lambda\right),\\
  \dot{\gamma}^a &=&\dot{y}^\alpha e_\alpha{}^a(\lambda) =
  \dot{y}^\alpha e_\alpha{}^a\left([y],\lambda\right),\\
\gamma^a(\lambda)&=& \int_0^\lambda d\lambda' \dot{y}^\alpha\label{7.17}
                     e_\alpha{}^a\left([y],\lambda\right) +x_o^a,\\
\gamma^a(0)&=&x_o^a,
\end{eqnarray}
where the brackets denote functional dependence on the $y's$. 
Notice that at the quantum level the local frames in (\ref{7.17}) will
be promoted to operators. If one describes the path in terms of the
intrinsic functions $y^\alpha(\lambda)$, the corresponding path in a given
system $\gamma^a$ will also be given by quantum operators, and therefore the
notion of point will only emerge in a semiclassical regime.

The tetrads defined allow to compute the
metric,
\begin{equation}
  g^{ab}\left([y],\lambda\right)=\eta^{\alpha\beta}
  e_\alpha{}^a\left([y],\lambda\right)  
e_\beta{}^b\left([y],\lambda\right)  =g^{ab}\left(\gamma(\lambda)\right).
\end{equation}

The assignment of frames $e_\alpha{}^a\left([y],\lambda\right)$ would
allow to identify that two different curves
have the same endpoints $\gamma^a_0(\lambda_f)=\gamma_1^a(\lambda_f)$. That implies,
\begin{equation}
  \int_0^{\lambda_f} d\lambda' \dot{y}_0^\alpha
  e_\alpha{}^a\left([y_0],\lambda'\right) = \int_0^{\lambda_f} d\lambda'
  \dot{y}_1^\alpha e_\alpha{}^a\left([y_1],\lambda'\right), 
\end{equation}
since the integrals reduce to evaluations at the endpoints of
$y's$. This is therefore the condition for two curves whose intrinsic
description is known, to have the same
endpoints. Since we are arriving at the same point with frames that
are parallel transported, they therefore may differ by a Lorentz
transformation, 
$e_\alpha{}^a\left([y_0],\lambda_f\right)=\Lambda_\alpha{}^\beta
e_\beta^a\left([y_1],\lambda_f\right)$ with $\Lambda_\alpha{}^\beta$ the
matrix of the Lorentz transformation. 

\section{Non-locality of the observable algebra}

Here we would like to analyze the non-locality of the observable
algebra in the linearized case. For that purpose we will define a
coordinate system in terms of reference paths for instance using
geodesics. In fact it is known that with the resulting Riemann normal
coordinates one may cover an arbitrarily large region of spacetime in
the linearized case \cite{Nesterov}. It is important to remark that
here we will not use the second order approximation for Riemann
normal coordinates. Let us first start by discussing how the linearized
theory emerges from the intrinsic formulation.

\subsection{From intrinsic gravity to linearized gravity}

Given such a coordinate system, 
we may now proceed as we did in section 5 and assign to each point $x$
in $U$ a spin connection, in the non-holonomic description given by
the tetrads $e^a_\alpha(\pi_R^x)$, 
\begin{equation}
  A_{\mu\alpha\beta}\left(x\right) =A_{\mu\alpha\beta}\left(F,\pi_R^x\right),
\end{equation}
where $\pi_R^x$ is the reference path defined above.

Analogously, 
\begin{equation}
  R_{\mu\nu\alpha\beta}\left(x\right) =R_{\mu\nu\alpha\beta}\left(\pi_R^x\right).
\end{equation}

In the linear approximation we can drop the second term in (\ref{5.8}),
\begin{equation}
  A_{\rho\gamma\nu}\left(x\right)=
  A_{\rho\gamma\nu}\left(\pi_R^x\right)=
\int_{s_i}^{s_f} ds \dot{y}^\alpha\left(s\right)
R_{\alpha\rho\gamma\nu}\left(\pi_R^{y(s)}\right),
\end{equation}
neglecting the correction of $A$ quadratic in $R$. Taking
into account that the $R$'s satisfy,
\begin{equation}
  R_{\alpha[\rho\gamma\nu]}\left(\pi\right)=0,
\end{equation}
one gets 
\begin{equation}
  A_{\rho\gamma\nu}\left(x\right)+
  A_{\nu\rho\gamma}\left(x\right)+
  A_{\gamma\nu\rho}\left(x\right)=0, \label{8.5}
\end{equation}
and from $\partial_{[\mu} R_{\alpha\beta]\gamma\nu}=0$, we get that,
\begin{equation}
  R_{\alpha\beta\gamma\nu} = \partial_\alpha A_{\beta\gamma\nu}-
\partial_\beta A_{\alpha \gamma\nu}.
\end{equation}
Finally, the symmetry
$R_{\alpha\beta\gamma\nu}=R_{\gamma\nu\alpha\beta}$ allows to define a
superpotential $h_{\rho\alpha}$,
\begin{equation}
  A_{\rho\alpha\beta}= h_{\rho\alpha,\beta}-h_{\rho\beta,\alpha}.
\end{equation}

Since the spin connections constructed satisfy (\ref{8.5}) for any
path, it follows that under a change of path
\begin{equation}
  A'_{\mu\alpha\beta}\left(x\right)  =
  A_{\mu\alpha\beta}\left(\pi'{}^x\right)
  =A_{\mu\alpha\beta}\left(x\right) + \Lambda_{\alpha\beta,\mu},
\end{equation}
just like in gauge theories, with,
\begin{equation}
  \Lambda_{\alpha\beta,\mu}+\Lambda_{\mu\alpha,\beta} +
\Lambda_{\beta\mu,\alpha}=0,
\end{equation}
and therefore define a vector $\xi_\alpha$, 
\begin{equation}
  \Lambda_{\alpha\beta} = \xi_{\alpha,\beta}-\xi_{\beta,\alpha},
\end{equation}
and
\begin{equation}
  \delta h_{\mu\alpha} =\xi_{\mu,\alpha} +\xi_{\alpha,\mu},
\end{equation}
and the components of the Riemann tensor are invariant under these
transformations. 

Recalling the relationship of spin connections with the tetrads in a
linear theory one gets,
\begin{equation}
  g_{\mu\nu} = \eta_{\mu\nu} +h_{\mu\nu}.
\end{equation}

These relations also hold for any assignment of reference paths which
satisfies the above mentioned conditions. That is: i) the end point
intrinsic coordinates defined by their total intrinsic displacement
along the parallel transported system of coordinates coincide with the
coordinates of the local chart in $V$; and ii) any portion of a
reference path is also a reference path.

It is important to emphasize that the intrinsic formulation of
linearized gravity also differs from the path dependent description of
gauge theories given in sections 2 and 3. Small differences depend on
the intrinsic description of the paths and would disappear if one
defines paths in a flat manifold and considers the linearized theory
as another gauge theory. For instance one would not even have an
equation (\ref{7.1}) in the case of ordinary (non-intrinsic) paths in
the flat background 
manifold.

One should however recall that only the intrinsic theory is given in
terms of physical observables. As we shall see a description in terms
of observables is always non-local. 

\subsection{Non-locality} 

In the case of gauge theories, like Yang--Mills theory, it is always possible to define local
gauge invariant observables, for instance ${\rm Tr}\left(F_{\alpha\beta}
  F^{\alpha\beta}\right)$. However, when gravity is included the
observables are always non-local. For example, a scalar field
$\phi\left(x\right)$ is not observable due to its dependence on
diffeomorphisms but $\phi\left(\pi\right)$ is since it refers to a
specific field at an intrinsically defined point and depends on a
non-ambiguous measuring procedure.

If one fixes paths, for instance using geodesics as in the previous section, the gauge is
completely fixed and the scalars are observable,
\begin{equation}
\phi\left(x\right) = \phi\left(\pi_R^x\right). 
\end{equation}

It is clear that in an eventual quantization, 
quantum fluctuations in the geometry throughout the path will change
the arrival point and therefore the value of the measured field. It is
difficult to do an explicit demonstration since the dependence enters
through the parallel transport whose expression in intrinsic
coordinates is not known explicitly. 

However, we can illustrate the dependence on the Riemann tensor by
considering the change of a scalar field when one changes the path.
For instance, if $x^\nu\left(\pi\right)$ are the Riemann normal coordinates of
the end point of a path $\pi$ that differs from $\pi_R^x$ by an
infinitesimal 
spatial deformation at $y$, an intermediate point of $\pi_R^x$, we
have that the coordinates of the end point (in the linearized case, to
keep things simple) change as, 
\begin{equation}
  x^\nu\left(\pi\right) = x^\nu -\frac{1}{2} \sigma^{\alpha\beta}
  R_{\alpha\beta \lambda}{}^\nu\left(\pi_R^y\right)
  \left(x_0-z\right)^\lambda. 
\end{equation}

\begin{figure}[h]
\includegraphics[height=5cm]{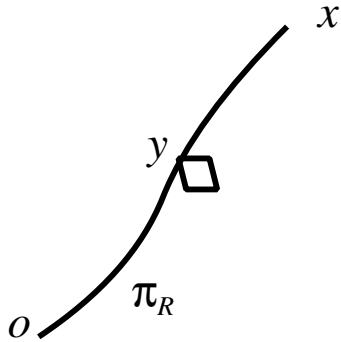}
\caption{The path $\pi_R$ with an infinitesimal deformation at the
  point $y$.}
\label{fig15}
\end{figure} 

Expressions like this suggest that scalar fields evaluated at
different points (or dependent on paths that end at different points)
will generically have non-vanishing Poisson brackets with the
Riemann tensor and therefore among themselves. This is what we mean by
non-locality. 

As we mentioned before, in this formulation the components of the Riemann tensor can be considered functions
of points given by the reference path,
\begin{equation}
  R_{\mu\nu\lambda\rho}\left(x\right) =  R_{\mu\nu\lambda\rho}\left(\pi_R^x\right),
\end{equation}
and in linearized gravity these quantities are gauge invariant and
therefore observables of the theory. We will show explicitly that the
non-locality of the theory emerges in the example of the linearized
case by noting that non-vanishing Poisson brackets between variables
at spatially separated points emerge. The observables will therefore
obey a non-local algebra.

The Poisson brackets between Riemann tensors of linearized gravity
were computed by De Witt some time ago. 
\begin{equation}
  \left[R_{\mu\nu\sigma\tau}\left(x\right),
R_{\alpha\beta\gamma\delta}\left(x'\right)\right]=\frac{i}{4} 
\left(\eta_{\mu\alpha} \eta_{\sigma\gamma}
  +\eta_{\mu\gamma}\eta_{\sigma\alpha}
  -\eta_{\mu\sigma}\eta_{\alpha\gamma}\right)\Delta\left(x,x'\right)_{,\nu,\tau,\beta,\delta}
+{\rm Permutations}\label{8.16}
\end{equation}
where $\Delta\left(x,x'\right)$ is the odd homogeneous propagator of
massless fields in flat space-times, and 
the permutations are the fifteen ones compatible with the
symmetries of the Riemann tensor. 

The variation of a scalar field under a change of path like we have
for a the spatial
deformation $\sigma^{ab}$ (we use Latin indices for spatial
components), is given by,
\begin{equation} 
\delta \phi\left(\pi,P\right)=\phi\left(\pi\right)
-\phi\left(\pi_R\right)=-\frac{1}{2}
\sigma^{kl}R_{kl\mu}{}^\nu\left(\pi_R^y\right)\left(x-y\right)^\mu \partial_\nu \phi\left(\pi_R\right).
\end{equation}
It can be easily checked from (\ref{8.16}) that it 
does not have vanishing Poisson bracket 
with the components of the Riemann tensor in $y$
where the deformation in the 
definition of $\delta\phi$ takes place. 
What we have shown for the variation $\delta\phi$ also holds for the field
itself if one considers the Poisson bracket with the Riemann tensor at 
a point along the path. Two fields themselves will also have
non-vanishing Poisson brackets as long as their paths intersect.

In contrast to what happens in ordinary field theory, gauge invariant
observables in the presence of gravity cannot be localized in well
defined regions of space-time and therefore one does not have a
definition of subsystems stemming from commuting
sub-algebras. Donnelly and Giddings have discussed gravitational
non-locality for a different set of gravitational observables in
references \cite{GiDo}.

It should be noted that at a classical level commuting subalgebras are
possible by considering observables dependent on non-intersecting paths.  At
a quantum level however, this is clearly impossible since given the
paths $\pi$ and $\pi'$ through their intrinsic description it is not
possible to know if they have or do not have intersections or common
parts when the geometry fluctuates and is not uniquely determined.

\section{The action in terms of path dependent fields}

Teitelboim \cite{teitelboim} was the first to note that the usual action of fields could
be expressed as an action of path dependent fields, by gauge fixing
using Fadeev--Popov terms. Although his proposal is very suggestive,
he does not present a proof of the equivalence with the ordinary
action. We will provide a proof for an arbitrary
Lagrangian, 
\begin{equation}
  \label{eq:20}
  {\cal L}\left(R_{\alpha\beta\gamma}{}^\rho\left(\pi\right),\phi\left(\pi\right),\psi\left(\pi\right)\right),
\end{equation}
where $\phi$ and $\psi$ are fields that could be scalar, vector or spinor.

The quantity ${\cal L}$ is a scalar and therefore is independent of
$\pi$. Let us recall that for scalar quantities $\pi$ only provides
the intrinsic description of the point in the manifold $M$ where it
is being evaluated. ${\cal L}$, given in terms of path dependent
quantities, does not refer to any local chart and therefore there is
no explicit reference to its invariance under diffeomorphisms.

If we describe $\pi$ through the path in the frame
parallel-transported from $o$, $x^\alpha(\lambda)$ with $u^\alpha =
dx^\alpha/d\lambda$ the action $S$ is given by $  S = \int {\cal L}
{\cal D} x$ with, 
\begin{equation}
{\cal D} x = \Pi_{\lambda,\alpha} dx^\alpha(\lambda) \delta(\pi-\pi'_R)\Delta_{FP}(\pi_R).
\end{equation}
We are considering a standard path integral integration, the product
on $\lambda$ represents the limit for $N$ going to infinity of the
product $\Pi_{i=1}^N$ for
partitions of the interval $[0,\lambda_f]$ in $N$ portions.
In the above expression $\pi_R$ is a reference path associated to each point of the
manifold $M$, $\delta(\pi-\pi_R)$ fixes a path for each point and
$\Delta_{FP}$ is the Fadeev--Popov determinant for that choice of
path. The Lagrangian ${\cal L}$ is a Lorentz scalar and takes the same
value for all paths $\pi$ that reach the end of $\pi_R$ and is
therefore independent of $\pi$. The choice of reference paths plays
the same role as a gauge fixing in an ordinary gauge theory.

Let us show the equivalence with the usual action in Riemann
normal coordinates in a neighborhood $U$ of a point $o_1$. We consider
paths $\pi^U$ and $\pi_R^U$  from $o_1$ to $P$ with $P$ and
arbitrary point in $U$. The path dependence will be restricted to the
region in which the normal coordinates are defined so we have that,
\begin{eqnarray}
\pi_R &=& \pi_{oR}^{o_1}\circ \pi_R^U,\\
\pi &=& \pi_{oR}^{o_1}\circ \pi^U,
\end{eqnarray}
and $\pi_{oR}^{o_1}$ a fixed reference path from $o$ to $o_1$. The
paths from $o_1$ to $P$, $\pi_R^U$ are geodesics and take the form
$x^\alpha=u^\alpha\lambda$.  If we define Riemann normal coordinates
associated with the geodesics centered at $o_1$, $z^a$, one can
identify $z^a=x^\alpha$ and the metric is locally flat at $o_1$.  Let
$\pi^U$ be a path from $o_1$ to $P$ given by $x_\pi^\alpha(\lambda)$ arbitrary such that
$x^\alpha(0)=0$. If $w^\alpha(\lambda)$ are the infinitesimal
displacements referred to the reference path that goes from
$x^\alpha_{\pi_R}(\lambda)$ to $x^\alpha_\pi(\lambda)$ and
$u^\alpha_\pi=dx^\alpha_\pi/d\lambda$, one has taking into account
(5.20) and (5.21), particularized to a geodesic path,
\begin{equation}
  \label{eq:22}
  u^\alpha_\pi(\lambda)=\Lambda^\alpha{}_\beta(\lambda) u^\beta+\frac{dw^\alpha(\lambda)}{d\lambda},
\end{equation}
with,
\begin{equation}
  \label{eq:23}
  \Lambda^\alpha{}_\beta(\lambda)=
\delta^\alpha_\beta+\Omega^\alpha_\beta(\lambda)=
\delta^\alpha_\beta+\int_0^\lambda d\lambda'
  R_{\gamma\delta}{}^\alpha{}_\beta(\lambda')
  w^\delta(\lambda')u^\gamma. 
\end{equation}

If we now impose the gauge conditions that say that the reference
paths are geodesics going from $o_1$ to $P$, $du^\alpha_\pi/d\lambda=d^2
x^\alpha_\pi/d\lambda^2=0$,
\begin{equation}
  \label{eq:24}
 0=\frac{d u^\alpha_\pi(\lambda)}{d\lambda}= R_{\gamma
   \delta}{}^\alpha{}_\beta(\lambda) u^\gamma
 w^\delta(\lambda) u^\beta+ \frac{d^2 w^\alpha(\lambda)}{d\lambda^2},
\end{equation}
where $u^\gamma, u^\beta$  are the constant vectors that define the
geodesic reference path $\pi_R$,
we get the equation that allows to compute the Fadeev--Popov
determinant. In order to do that we note that, integrating (\ref{eq:22}),
\begin{equation}
  \label{eq:25}
  \delta\left(x^\alpha_\pi(\lambda)-x^\alpha_R(\lambda)\right)=\delta
  \left( \int_0^\lambda \Omega^\alpha{}_\beta(\lambda')
    u^\beta d\lambda'+w^\alpha(\lambda)\right)
\end{equation}
where $w^\alpha(\lambda=0)=w^\alpha(\lambda=\lambda_f)=0$ since both 
paths go from $o_1$ to $P$. We also recall that,
\begin{equation}
  \label{eq:26}
  \delta\left(\pi-\pi_R\right)=\Pi_{\lambda,\alpha}
  \delta\left(x^\alpha_\pi(\lambda)-\lambda u^\alpha\right).
\end{equation}

Let us note that the above expression can be written as,
\begin{equation}
  \label{eq:27}
  \delta\left(x^\alpha_\pi(\lambda)-x^\alpha_R(\lambda)\right)=\delta\left(M^{\alpha(\lambda)}{}_{\delta
    (\mu)} w^{\delta(\mu)}\right).
\end{equation}
where $\alpha$ and $\delta$ are Lorentz indices and $(\lambda)$ and
$(\mu)$ continuous indices that are integrated from $0$ to $\lambda$
when repeated and $w^{\delta(\mu)}\equiv w^\delta(\mu)$.  The quantity
$M^{\alpha(\lambda)}{}_{\delta (\mu)}$ can be computed by first
integrating (\ref{eq:24}) for $w^\alpha$ with boundary conditions that
vanish for $\lambda=0$ and $\lambda=\lambda_f$,
i.e. $w^\alpha(0)=w^\alpha(\lambda_f)=0$. With this, in a sufficiently
small region $U$, allowing the second order approximation for Riemann
coordinates, we have that,
\begin{eqnarray}
  \label{eq:29}
  M^{\alpha(\lambda)}{}_{\beta (\mu)}&=&
                                   \delta^\alpha_\beta\delta(\lambda-\mu)+R_{\gamma
                                   \beta}{}^\alpha{}_\delta(0)u^\gamma
                                   u^\delta\nonumber\\
&&\times\left[\int_0^\lambda d\lambda'\int_0^{\lambda'}d\lambda"
                                   \delta(\lambda"-\mu)-\frac{\lambda}{\lambda_f}
                                   \int_0^{\lambda_f}d\lambda'
                                   \int_0^{\lambda'}d\lambda"
                                   \delta(\lambda"-\mu)\right]\nonumber\\
&=&
    \delta^\alpha_\beta\delta(\lambda-\mu)+R_{\gamma\beta}{}^\alpha{}_\delta
    u^\gamma u^\delta
\left[\int_0^\lambda \Theta(\lambda'-\mu)d\lambda'
    -\frac{\lambda}{\lambda_f}\int_0^{\lambda_f} d\lambda'
    \Theta(\lambda'-\mu)\right]\nonumber\\
&=& \delta^\alpha_\beta\delta(\lambda-\mu)+R_{\gamma\beta}{}^\alpha{}_\delta
    u^\gamma u^\delta\left[(\lambda-\mu)-\frac{\lambda}{\lambda_f}(\lambda_f-\mu)\right].
\end{eqnarray}

We now recall that $\delta(L x )=\delta(x)/{\rm det}(L)$ if $L$ is a
non-singular matrix and $x$ a vector. Therefore the right hand side
of (\ref{eq:27}) can be rewritten involving the determinant of
$M$. This is, by definition, the Fadeev--Popov determinant.

In order to compute the determinant we use that,
\begin{equation}
  \label{eq:30}
  {\rm det}(I+L)=\exp\left({\rm Tr} \ln(I+L)\right),
\end{equation}
where $I$ is the identity matrix and $L$ is a matrix of norm smaller
that one.  This holds in  this case since $R \lambda_f^2\ll 1$. 
Recalling that we are working up to second order in the Riemann
coordinates only, we can take $\ln(I+L)\sim L$ and keep only the first
order in the exponential. Therefore the determinant is given by
$1+{\rm Tr}(L)$. Therefore,
\begin{eqnarray}
  \label{eq:32}
  {\rm det}(I+L)&=&1 -\int_0^{\lambda_f} R_{\gamma
  \beta}{}^\beta{}_\delta u^\gamma u^\delta (\lambda_f
                    -\lambda)\frac{\lambda}{\lambda_f}d\lambda\nonumber\\
&=& 1-R_{\gamma
  \beta}{}^\beta{}_\delta u^\gamma u^\delta \left.\left(\frac{\lambda_f
    \lambda^2}{2}
    -\frac{\lambda^3}{3}\right)\right\vert_0^{\lambda_f}\frac{1}{\lambda_f}\nonumber\\
&=& 1-R_{\gamma
  \beta}{}^\beta{}_\delta u^\gamma u^\delta \frac{\lambda_f^2}{6} =
1 -R_{cb}{}^{bd}\left(\pi_{oR}^{o_1}\right)\frac{z^cz^d}{6},
\end{eqnarray}
with $z^a=u^a\lambda_f$. Recall that in normal coordinates we have that,
\begin{equation}
  \label{eq:33}
  g_{mn}=\eta_{mn}-\frac{1}{3} R_{m a n b}z^a z^b,
\end{equation}
so for the
determinant of the metric we have that,
\begin{eqnarray}
  \label{eq:34}
  \sqrt{-g}&=& \sqrt{1
-\frac{1}{3} \eta^{mn}R_{m a n b}u^a
               u^b\lambda_f^2}\nonumber\\
&=&1 -\frac{1}{6} \eta^{m n}R_{m a n b}z^az^b.
\end{eqnarray}

And therefore,
\begin{equation}
  \label{eq:35}
  {\cal D} x = \Pi_{\alpha,\lambda} dx^\alpha_\pi(\lambda)
  \delta(\pi-\pi_R)\Delta_{FP}=\Pi_a dz^a \sqrt{-g},
\end{equation}
and we recover the Einstein--Hilbert action in the coordinate system defined by
the $\pi_R$.

\section{Some comments about the canonical formulation}

To get some idea of the issues involved in a canonical quantization of 
a framework like the one presented, let us consider the particular
case of a scalar field in a curved space-time and study its canonical
formulation.

A path dependent scalar field $\phi(\pi)$ is such that
$\phi(\pi')=\phi(\pi)$ if $\pi'=\gamma\cdot \pi$ with $\gamma$ a
closed loop. Indeed, its description is frame independent 
and the path only fulfills the role of identifying the point
in $M$ where the field is evaluated without introducing coordinate
systems. 

The equation of motion for a massless scalar field in a Riemannian
manifold is, 
\begin{equation}
  \label{eq:42}
  {\eta}^{\alpha\beta} D_\alpha D_\beta \phi(\pi)=0,
\end{equation}
and follows from the action,
\begin{equation}
  \label{eq:43}
S =\frac{1}{2} \int D u_\pi \Delta_{FP}(\pi) \delta(\pi'-\pi)
\eta^{\alpha\beta} D_\alpha \phi(\pi') D_\beta \phi(\pi'),
 \end{equation}
following the ideas of previous sections.

\begin{figure}[h]
\includegraphics[height=5cm]{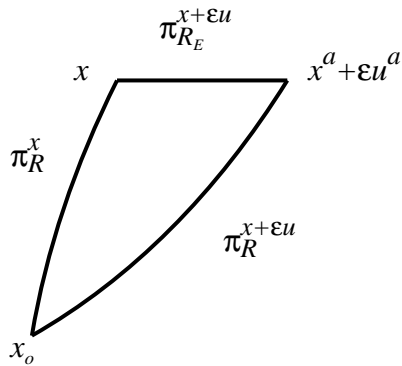}
\caption{The path in the Mandesltam derivative}
\label{fig17}
\end{figure}

Let $\pi=\pi_R$ be an arbitrary assignment of reference paths.
By definition of the Mandelstam derivative we have that (see figures
12 and \ref{fig17}),
\begin{equation}
  \label{eq:47}
  \left(1+\epsilon u^\alpha D_\alpha\right)
  \phi\left(\pi_R^x\right)=\phi\left(\pi_{R_E}^{x+\epsilon u}\right),
\end{equation}
where $\pi_{R_E}$ represents the extended path.

Since the scalar field in $x+\epsilon u$ takes the same value for any
path with that final point the covariant derivative becomes an
ordinary one and we have that,
\begin{equation}
  \label{eq:48}
  u^\alpha D_\alpha \phi\left(\pi_R\right)=u^\alpha
  e_\alpha{}^a\left(\pi_R\right) \left.\left(\partial_a
      \phi\left(\pi_R\right)\right)\right\vert_{\pi_R^x}=
  u^a \partial_a \phi\left(\pi_R^x\right), 
\end{equation}
where $e_\alpha^a$ is the frame transported along $\pi_R$,
\begin{equation}
  D_\alpha \phi\left(\pi_R\right)=
  e_\alpha{}^a\left(\pi_R\right) \partial_a \phi\left(\pi_R^x\right).   
\end{equation}

As a consequence, the action becomes the ordinary one,
\begin{eqnarray}
  S &=& -\frac{1}{2} \int {\cal D} u_{\pi_R}
        \Delta_{FP}\left(\pi_R\right) \delta\left(\pi'-\pi_R\right)
        \eta^{\alpha\beta} e_\alpha{}^a\left(\pi_R\right)
        e_\beta{}^b\left(\pi_R\right) \partial_a
        \phi\left(x\left(\pi_R\right)\right)
\partial_b \phi\left(x\left(\pi_R\right)\right),\nonumber\\
&=& -\frac{1}{2} \int dx \sqrt{-g} g^{ab}(x) \partial_a
    \phi \partial_b \phi.
  \end{eqnarray}

Its equations of motion are,
\begin{equation}
  g^{ab} \nabla_a \nabla_b \phi(x) =   
  \eta^{\alpha\beta} e_\alpha{}^a e_\beta{}^b \nabla_a \nabla_b \phi =
  \eta^{\alpha\beta}D_\alpha D_\beta \phi\left(\pi_R^x\right)=0,
\end{equation}
where we have used equation (\ref{7.5}). 

Let us proceed to the canonical formulation. In the first place we
note that although $\phi(\pi)$ is independent of the path $\pi$ that
arrives at the point $x$, its canonical conjugate momentum depends of
the notion of time used and therefore of the frame transported to $x$
along $\pi$. It should be pointed out that the canonical framework is
not well suited for the intrinsic formulation since it assumes that
the surface can be foliated and that the topology is fixed. These are
two hypotheses that are not natural in the intrinsic approach. It
will, however, allow us to carry out a first approach towards
quantization. 

\begin{figure}[h]
\includegraphics[height=5cm]{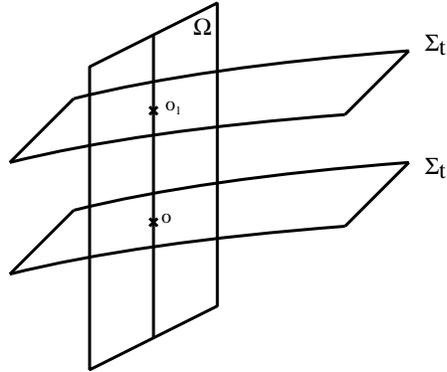}
\caption{The foliation of the manifold.}
\label{fig18}
\end{figure}

We can start with a manifold $M=\Sigma\times R$ with coordinates
adapted and introduce a geometry in $M$ \`a la ADM, for example.  In
order to introduce intrinsic reference paths arriving to each point of
$M$ adapted to a foliation $\Sigma_t$ (see figure 14), we introduce a platform through $o$,
a three dimensional hypersurface $\Omega$ such that
$\Sigma_t\cap \Omega$ is two dimensional. We also introduce a congruence
of curves in $\Sigma_t$ parameterized by $t,u,v,w$ such that $\gamma^a\left(t,u,v,w_0=0\right)$ are
points on $\Sigma_t \cap \Omega$ and that
$\gamma^a\left(t=t_0,0,0,0\right)$ is the origin of the intrinsic
description. Given a point $t_1, x_1$ with
$x_1^i=\gamma^i\left(t_1,u_1,v_1,w_1\right)$ in $\Sigma_{t_1}$ we
define reference paths $\pi_R$ starting in $o$, $\gamma^a(t,0,0,0)$ with
$\gamma^a(t_1,0,0,0)=o_1$. From $o_1$ we go to
$x'_1{}^i=\gamma^i(t_1,u_1,v_1,0)$ through the path
$\gamma^a(t_1,\lambda u_1,\lambda v_1,0)$ with $\lambda\in [0,1]$ and
through $\gamma^a(t_1,u_1,v_1,w)$ to $\gamma^a(t_1,u_1,v_1,w_1)$. Its
intrinsic description will depend on the geometry. If we use a $3+1$
ADM notation, we have that,
\begin{eqnarray}
  g_{ij} &=& {}^4g_{ij},\qquad N =
             \left(\sqrt{-{}^4g^{00}}\right)^{-1},\qquad N_i={}^4
             g_{0i},\\
g^{ij}g_{jk} &=& \delta^j_k,\qquad {}^4 g_{00} =
  -\left(N^2-N^iN_i\right), \qquad N^i=g^{ij}N_j,\\
{}^4g^{0i}&=&\frac{N^i}{N^2}, \qquad {}^4 g^{00}=-\frac{1}{N^2},\qquad
{}^4 g^{ij} = g^{ij}-\frac{N^i N^j}{N^2},\\
{\rm det}{}^4g&=& - N {\rm det}g,\qquad \sqrt{-{}^4 g} = N \sqrt{g},\\
n_\mu &=& -N \delta_\mu^0, \qquad {}^4 g^{\mu\nu}n_\mu n_\nu=-1.
\end{eqnarray}

Recalling that the action for $\phi(x)=\phi\left(\pi_R^x\right)$ is,
\begin{eqnarray}
  S&=& -\frac{1}{2} \int dx \sqrt{-g} g^{AB} \partial_A \phi \partial_B
  \phi,\nonumber\\
&=&-\frac{1}{2} \int dx N\sqrt{g} \left[-\frac{1}{N^2}
    \left(\partial_0 \phi\right)^2+2 \frac{N^i}{N^2} \partial_0
    \phi \partial_i \phi
    +\left(g^{ij}-\frac{N^iN^j}{N^2}\right) \partial_i \phi \partial_j \phi\right], 
\end{eqnarray}
the canonical momentum is
\begin{equation}
P_\phi = \frac{\sqrt{g}}{N} \partial_0 \phi
-\frac{N^i}{N}\sqrt{g} \partial_i \phi.  \label{11.14}
\end{equation}

We can then proceed to do the Legendre transform and obtain the
Hamiltonian,
\begin{eqnarray}
  {\cal H} &=& P_\phi \partial_0 \phi -{\cal L}\nonumber\\
&=& \frac{N P_\phi^2}{\sqrt{g}}+P_\phi N^i \partial_i \phi
-\frac{1}{2} \frac{\sqrt{g}}{N}\left(\frac{N
    P_\phi}{\sqrt{g}}+N^i\partial_i \phi\right)\nonumber\\
&&+\frac{N^i\sqrt{g}}{N} \partial_i \phi \left(\frac{N
   P_\phi}{\sqrt{g}}+N^i\partial_i\phi\right)
+\frac{1}{2}\left(g^{ij} -\frac{N^i N^j}{N^2}\right)\partial_i
   \phi \partial_j \phi \sqrt{g}\nonumber\\
&=& \frac{N P_\phi^2}{2\sqrt{g}} +P_\phi N^i\partial_i \phi+
\frac{1}{2}\left( g^{ij}\partial_i \phi \partial_j \phi\right) N \sqrt{g}.  
\end{eqnarray}

From it we get the equations of motion
\begin{eqnarray}
  \partial_0 \phi &=& \frac{N P_\phi}{\sqrt{g}} + N^i\partial_i \phi,\label{11.16}\\
\partial_0 P_\phi &=& -\partial_i \left(P_\phi N^i\right)-\partial_i
                      \left(g^{ij}\partial_j \phi N \sqrt{g}\right).
  \end{eqnarray}

The Poisson brackets are,
\begin{eqnarray}
  \left\{ \phi(x),P_\phi(y)\right\}_t &=& 
  \left\{ \phi(\pi^x),P_\phi(\pi^y)\right\}_t =\delta(x,y),\\
  \left\{ \phi(x),\phi(y)\right\}_t &=& 
  \left\{ P_\phi(x),P_\phi(y)\right\}_t =0.
  \end{eqnarray}

From (\ref{11.14},\ref{11.16}) we get the brackets of the time
derivatives,
\begin{eqnarray}
\label{10.20}  \left\{\phi(x),\partial_0 \phi(y)\right\} &=& \frac{N}{\sqrt{g}}
                                                \delta(x,y),\\
\left\{ \partial_0 \phi(x), P_\phi(y)\right\} &=& N^i\partial_i \delta(x,y),
\end{eqnarray}
where $\partial_0$ is the derivative with respect to the parameter of
the foliation $\Sigma_t$. 

So we see that fixing the reference path has led us to the traditional 
canonical formulation of scalar field. 
But we really are interested in the Poisson brackets for arbitrary
paths described intrinsically. Let us first consider paths that start
from $o$ in $\Sigma_t$ and have the same end point than that of
$\pi^x$. In order to do that we will use the technique of going from
paths in coordinate systems to intrinsic paths and vice-versa. It will
allow us to recognize paths that end in $x$. 

Let $\pi$ given by $y^\alpha(\lambda)$ that corresponds to
$\gamma^a(\lambda)$ with $\gamma^\mu(x)=x^a$, that is,
\begin{equation}
  \int_0^{\lambda_f} d\lambda \dot{y}^\alpha(\lambda) e_\alpha{}^a\left([y],\lambda\right)=x^a,  
\end{equation}
and $\pi'$ given by $y'{}^\alpha(\lambda)$,
\begin{equation}
  \int_0^{\lambda_f} d\lambda \dot{y'}^\alpha(\lambda) e_\alpha{}^a\left([y'],\lambda\right)=z^a.  
\end{equation}

The Poisson brackets satisfy 
\begin{equation}
  \label{eq:21}
  \left\{ \phi(\pi),P_\phi(\pi')\right\} = \delta^3\left(\gamma^a(\lambda_f),\gamma'{}^a(\lambda_f)\right)=\delta^3(x^a,z^a),
\end{equation}
with 
\begin{equation}
\label{10.25}\gamma^a(\lambda)=  \int_0^\lambda d\lambda_1 \dot{y}^\alpha(\lambda_1) e_\alpha{}^a\left([y],\lambda_1\right).
\end{equation}

The advantage of this kind of relation is that it is easily
generalizable to the case of quantum gravity where the $e_\alpha{}^a$
are operators. 

If we consider $\pi$ extended to the future region defined
along the time component of the local basis
$e_\alpha{}^a\left([y],1\right)$, we have
that,
\begin{eqnarray}
\label{10.26}D_0 \phi(\pi) &=& e_0{}^a \partial_a \phi(\pi),\\
\label{10.27} \left\{ D_0 \phi(\pi), \phi(\pi')\right\}&=& -\frac{e_0{}^0
                                             N}{\sqrt{g}} \delta\left(\gamma^a(\lambda_f),\gamma'{}^a(\lambda_f)\right).  
\end{eqnarray}

The timelike Mandelstam derivative extends the path $\pi$ along the
zeroth component of the parallel transported frame and is given by
(\ref{10.26}). From it and (\ref{10.20}) we get (\ref{10.27}). 

Notice that if one were to quantize the gravitational field that
equation ({\ref{10.27}) taking into account the relation of the
  intrinsic and space-time coordinates (\ref{10.25}) would become an
  operatorial identity. In particular the arguments of the Dirac delta
in (\ref{10.27}) become operatorial. This will require further study for a complete
canonical quantization. As we mentioned, the canonical approach is not
the most natural in this context and other approaches that implement
directly the algebra of Dirac observables might be preferred.

\section{Concluding remarks}

We presented an intrinsic framework for the formulation of
gravitational theories including general relativity in terms of paths.
We solved the problem of defining what is a space-time point, that was
problematic in the original proposal on the subject by Mandelstam.  The
relation of the fields for two paths that arrive at the same point is
now under control.

In the intrinsic description of gravity a physical point is given by
the equivalence class of paths that differ by loops that may be
defined by the repeated action of the loop derivative. In an eventual
quantum theory, a fluctuation of the geometry in any region of
space-time will change that equivalence class, that is, some of the
paths that led to that point will fail to arrive to it. This will
induce fluctuations in the points that must be considered as emergent
objects of an underlying structure of paths. The fluctuations of the
space-time points will be more important in a region where quantum
effects are expected to be large, like near where black holes have
their classical singularities.  Close to a region with big quantum
fluctuations the fields will stop being local, in particular scalar
fields associated to nearby points will not commute, irrespective of
the separation being space-like or time-like. Note that the
non-locality is also in time, which makes the causal structure of
events become fuzzy. One of the remaining questions is whether is
relevant in the context of black hole evaporation.

The intrinsic description naturally operates with space-time paths. 
However, even if
one considers spatial paths one could end up in points that are in the
future of where one started. This will require special care at the
time of quantization, as was already observed by Mandelstam.

The whole construction is locally Lorentz invariant but there may be a
distortion of the invariance, unrelated to the ones due to
granular descriptions of space-time, due to the fluctuation of the
points. 
Further studies of the quantization are needed to understand the
non-local effects induced by time-like paths. In a forthcoming paper
we will discuss the Poisson algebra of path dependent fields including
gravity and its quantization.

\section{Acknowledgements}
We wish to thank Aureliano Skirzewski and Rafael Porto for discussions
and especially Miguel Campiglia and Saeed Rastgoo for
reading the manuscript and providing useful advice and comments.
This work was supported in part by Grant
No. NSF-PHY-1603630, funds of the Hearne Institute for Theoretical
Physics, CCT-LSU, and Pedeciba and Fondo Clemente Estable
FCE\_1\_2014\_1\_103803.

\section*{Appendix} 

To understand the effects of curvature on finite closed paths one
needs to take into account that for the infinitesimal generator or the
loop derivative to close a loop after going along a path $\epsilon u
\epsilon w \epsilon \overline{u}_{||}$ with $u,v$ unit vectors, instead
of traversing $\overline{w}_{||}$ one needs to go along a different
path, which we call $\overline{w}^{(1)}$. We introduce here the notation
$u_{||}$ to emphasize that it is at a different point (and referred to
a different frame) than
$u$ since this will be important in the calculation. To compute it, we consider
normal coordinates around an arbitrary point of the manifold $o'$ that
can be considered at the end of a path $\pi_o^{o'}$. The geodesics emanating from
$o'$ are given in normal coordinates by $x^\mu(s)=v^\mu s$ (the
$x^\mu$ are the normal coordinates and $v\mu$ are constants). The metric at $o'$ is $\eta$ and,
the Christoffel symbols and metric nearby are given in normal
coordinates in terms of the Riemann tensor computed at $o'$ as,
\begin{eqnarray}
  \Gamma^\mu_{\alpha\beta} &=&
  -\frac{1}{3}\left(R^\mu{}_{\alpha\beta\gamma}+R^\mu{}_{\beta\alpha\gamma}\right)
  x^\gamma,\\
g_{\mu\nu}(x)&=& \eta_{\mu\nu} -\frac{1}{3}
R_{\mu\alpha\nu\beta}x^\alpha x^\beta.
\end{eqnarray}

\begin{figure}[h]
\includegraphics[height=8cm]{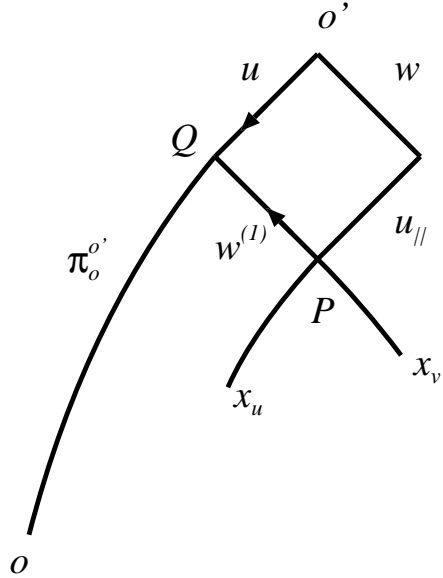}
\caption{The geodesics discussed in the appendix.}
\label{fig10}
\end{figure} 

We compute $u(s)$ by transporting $u$ along $\epsilon w$,
\begin{eqnarray}
  w^\alpha u^\beta{}_{;\alpha}&=&w^\alpha\left(\partial_\alpha
    u^\beta+\Gamma^\beta_{\alpha\rho} u^\rho\right)=0,\\
w^\alpha\partial_\alpha u^\beta= \frac{du^\beta}{ds} &=&
-w^\alpha \Gamma^\beta_{\alpha\rho}u^\rho=\frac{1}{3} R^\beta{}_{\alpha
  \rho \gamma} w^\alpha u^\rho w^\gamma s,\\
\frac{d^2 u^\beta}{d s^2} &=&\frac{1}{3}
R^\beta{}_{\alpha\rho\gamma}w^\alpha u^\rho w^\gamma.
\end{eqnarray}
Therefore,
\begin{eqnarray}
  u^\beta(s) &=& u^\beta(0)+\frac{1}{6} R^\beta{}_{\alpha\rho\gamma}
  w^\alpha u^\rho w^\gamma s^2,\\
  u^\beta_{||} &=& u^\beta(0)+\frac{1}{6} R^\beta{}_{\alpha\rho\gamma}
                   w^\alpha u^\rho w^\gamma \epsilon^2,\\
\end{eqnarray}
and transporting $w$ along $\epsilon u$ we get,
\begin{equation}
  w^\beta_{||} = w^\beta(0)+\frac{1}{6} R^\beta{}_{\alpha\rho\gamma}
  u^\alpha w^\rho u^\gamma \epsilon^2,
\end{equation}
where $u^\beta_{||}$ is $u$ parallel transported along $\epsilon w$
and is shown in figure (\ref{fig10}) and $w^\beta_{||}$, not shown in
the figure is $w$ parallel transported along $\epsilon u$. The vectors
$u,w, u_{||}, w_{||}$ do not form a closed loop. To close it we need
to compute $w^{(1)}$ which differs by terms of order $\epsilon^3$ 
from $w_{||}$.

In Riemann coordinates an arbitrary geodesic not necessarily going
through $o$ is given by, 
\begin{equation}
  x^\mu(s)=v_0^\mu+v_1^\mu s +\frac{1}{3}
  R^\mu{}_{\alpha\beta\rho}v_0^\rho v_1^\alpha v_1^\beta s^2,
\end{equation}
where $v_0^\mu$ are the coordinates of a point $v_0$ and $v_1^\mu$ the
tangent to the geodesic at the same point.

Let us consider the geodesic $x^\mu$ in figure 15  and let us determine
the coordinates of the point $P$, end point of $u_{||}$, and $Q$,
end point of $u$,  
\begin{equation}
  x_P^\mu=\epsilon w^\mu -\epsilon u^\mu +
  \frac{\epsilon^3}{6} R_{\alpha\rho\beta}{}^\mu u^\alpha w^\beta w^\rho 
  -\frac{\epsilon^3}{3} R_{\beta\rho\alpha}{}^\mu w^\rho u^\alpha u^\beta,
\end{equation}
and therefore $x_P-x_Q$ is given by,
\begin{equation}
x_P^\mu-x_Q^\mu= \epsilon w^\mu 
 +\frac{\epsilon^3}{6} R_{\alpha\rho\beta}{}^\mu u^\alpha  w^\rho w^\beta
  +\frac{\epsilon^3}{3} R_{\beta\rho\alpha}{}^\mu w^\rho u^\alpha
  u^\beta.
\end{equation}
Therefore the components of the vector $w^{(1)\mu}$ at the point $Q$
are given by, 
\begin{equation}
  \epsilon w^{(1)\mu}= \epsilon w^\mu
  +\frac{\epsilon^3}{6} R_{\alpha\rho\beta}{}^\mu u^\alpha w^\rho w^\beta 
  +
  \frac{\epsilon^3}{3} R_{\alpha\rho\beta}{}^\mu u^\beta u^\alpha w^\rho 
  ,
\end{equation}
and notice that $w^{(1)}$ is not a unit vector.  It is also useful to
compute the components of $w^{(1)}$ at the point $P$. 
In the intrinsic notation we
need to write $w^{(1)}$ in the parallel transported basis to $P$ given
by, 
\begin{equation}
  e_\alpha{}^\mu\left(P\right) = 
\delta_\alpha{}^\mu +\frac{1}{3} R_{\rho \alpha\beta}{}^\mu u^\beta
w^\rho \epsilon^2+\frac{1}{6} R_{\rho \alpha\beta}{}^\mu u^\beta
u^\rho \epsilon^2.
\end{equation}
Therefore $\epsilon w^{(1)}$ in intrinsic notation takes the form,
\begin{equation}
  \epsilon w^{(1)\alpha} =-\epsilon w^\alpha -\frac{\epsilon^3}{2}
  R_{\gamma \rho \beta}{}^\alpha u^\beta u^\gamma w^\rho -
\frac{\epsilon^3}{2} R_{\gamma\rho\beta}{}^\alpha u^\gamma w^\rho w^\beta.
\end{equation}

\end{document}